\documentclass[a4paper,12pt]{article}
\ifx\TeXtures\undefined
\usepackage[active]{srcltx} 
\usepackage[all]{xy}

\usepackage{hyperref}

\hypersetup{ 
colorlinks=true, 
breaklinks=true, 
urlcolor= black, 
linkcolor= black, 
bookmarksopen=true,
pdftitle={}, pdfauthor={}, pdfsubject={}
}

\usepackage[psamsfonts]{amsfonts}%
\usepackage{amsmath}%
\else
\usepackage{amsmath}%
\usepackage{amsfonts}%
\usepackage{times}
\fi
\usepackage{amsthm} 
\usepackage{epsfig}         
\usepackage[utf8]{inputenc}
\usepackage[T1]{fontenc}
\usepackage[english]{babel}
\usepackage{ae,icomma} 
\usepackage{amssymb,amscd,verbatim}
\usepackage{appendix}
\newcommand{\DATUM}{20-07-2025}              
\newcommand{\change}
{{\marginpar{\#}}}        
\newcommand{\comma}{\: ,}              
\newcommand{\period}{\: .}             
\newcommand{\cC}{{\cal C}}
\newcommand{\cD}{{\cal D}}
\newcommand{\cE}{{\cal E}}

\newcommand{\cG}{{\cal G}}

\newcommand{\cN}{{\cal N}}         
\newcommand{\cO}{{\cal O}}         

\newcommand{\cR}{{\cal R}}
\newcommand{\cS}{{\cal S}}

\newcommand{\cU}{{\cal U}}
\newcommand{\cV}{{\cal V}}
\newcommand{\cW}{{\cal W}}

\newcommand{\cY}{{\cal Y}}

\newcommand{\field}[1]{\mathbb{#1}}
\newcommand{\R}{\field{R}}            
\newcommand{\N}{\field{N}}            
\newcommand{\C}{\field{C}}            

\newcommand{\ut}{{\underline t}}
\newcommand{\ux}{{\underline x}}
\newcommand{\uz}{{\underline z}}

\newcommand{\uxi}{{\underline\xi}}
\newcommand{\rL}{{\rm L}}                 
\newcommand{\rC}{{\rm C}}

\newcommand{\rW}{{\rm W}} 
\newcommand{\rR}{{\rm R}} 
\newcommand{\rU}{{\rm U}} 
\newcommand{\rrS}{{\rm S}} 
\newcommand{\rx}{{\rm x}}

\newcommand{\mfz}{\mathfrak{z}} 
 
\newcommand{\umfz}{\underline{\mathfrak{z}}}

\newcommand{\cirS}{\mathop{\bigcirc\kern -.73em {\scriptstyle{\rm S}}}}

\newcommand{\cqfd}{\phantom{blablabla}\hfill\qed\newline} 
\newtheorem{theorem}{Theorem}[section]         
\newtheorem{lemma}[theorem]{Lemma}             
\newtheorem{definition}[theorem]{Definition}   
\newtheorem{remark}[theorem]{Remark}           
\newtheorem{proposition}[theorem]{Proposition} 

\theoremstyle{plain}

\newcommand{\donne}{\mapsto}
\newcommand{\dans}{\longrightarrow}

\newcommand{\Pf}{\vspace*{-2mm}{\bf Proof:}\, }
\newcommand{\Pfof}[1]{{\bf Proof of #1:}\, }

\renewcommand{\theequation}{\thesection.\arabic{equation}}

\newcommand{\ncg}{{[\hskip-.7mm [}}
\newcommand{\ncd}{{]\hskip-.7mm ]}}

\begin{document}

\setcounter{section}{0} 

\title{Regularity of the (N-1)-particle \\
electronic reduced density matrix \\
for molecules with fixed nuclei and N electrons.}
\author{
{\bf Thierry Jecko}\\
AGM, UMR 8088 du CNRS, site de Saint Martin,\\
2 avenue Adolphe Chauvin,\\
F-95000 Cergy-Pontoise, France. \\
e-mail: jecko@math.cnrs.fr\\
web: http://jecko.perso.math.cnrs.fr/index.html
\\
{\bf Camille Noûs\footnote{The fictitious author Camille Noûs embodies the collegial nature of the present work, as a reminder that science proceeds from disputatio and that the building and dissemination of knowledge are intrinsically selfless, collaborative and open. For details, see:
\[``\ https://www.cogitamus.fr/indexen.html  \ ''\period\]
}}\\
Laboratoire Cogitamus\\
e-mail: camille.nous@cogitamus.fr\\
web: https://www.cogitamus.fr/
}
\date{\DATUM}
\maketitle

This is a preprint version of the paper published in Letters in Mathematical Physics at:\\

\centerline{`` https://doi.org/10.1007/s11005-025-01975-4 ''.}

\begin{abstract}
We consider an electronic bound state of the usual, non-relativistic, molecular Hamiltonian with Coulomb interactions, fixed nuclei, and $N$ electrons ($N>1$). Near appropriate electronic collisions, we determine the regularity of the $(N-1)$-particle electronic reduced density matrix. 
\vspace{2mm}

\end{abstract}






\section{Introduction.}
\label{intro}
\setcounter{equation}{0}

Motivated by the Density Functional Theory (DFT) (cf. \cite{e,lise}), we aim, as in \cite{jn}, to study the regulartity of some specific object that is associated by the DFT to an electronic bound state of a molecular, Coulombic Hamiltonian with fixed nuclei and $N$ electrons (with $N>1$). In the past, many results proved the real analyticity on large domains of the electronic (reduced) density matrices, that are associated to such a bound state: see \cite{fhhs1,fhhs2,hs1,j1,j2}. Are these results optimal? More precisely, is the real analyticity of these matrices valid on a larger domain? We refer to \cite{j2} for a review on the proofs of the real analyticity of these matrices and to \cite{jn} for a review of results on their optimality and of strategies to treat it. \\
In \cite{jn}, for the $(N-1)$-particle reduced density matrix (defined in \eqref{eq:gamma-densité}) (actually, the conjugate was considered but this does not affect the results), it was proven that the real analyticity breaks down near some points of the boundary of the domain on which its real analyticity is known. It was even shown that the $(N-1)$-particle reduced density matrix cannot be smooth near such points and one has some information on its regularity there. \\
In the present paper, we improve the results in \cite{jn} in two directions: we enlarge the set of points where the non-smoothness of the $(N-1)$-particle reduced density matrix is proven and, more importantly, we determine its regularity (in the sense defined below) near those points. In particular, we address the problem of a ``fifth order cusp'' for the $(N-1)$-particle reduced density matrix, that appears in the literature in Chemistry (cf. \cite{c1,c2}). See also \cite{he}. \\
We point out that the lack of smoothness of density matrices is studied with another point of view in \cite{hs2,so1,so2}. \\
As in \cite{jn} and for the same reason, our method allows us to treat the $(N-1)$-particle reduced density matrix only. We believe that the extension of the present results to other reduced density matrices or reduced densities is a quite involved task. 

Let us first present our framework. We consider a molecule with $N$ moving electrons, with $N>1$, and $L$ fixed nuclei, with $L\geq 1$ (according to Born-Oppenheimer idealization). The $L$ distinct vectors $R_1, \cdots , R_L\in\R^3$ represent the positions of the nuclei. The positions of the electrons are given by $x_1, \cdots , x_N\in\R^3$. The charges of the nuclei are respectively given by the positive $Z_1, \cdots , Z_L$ and the electronic charge is set to $-1$. The Hamiltonian of the electronic system is 
\begin{eqnarray}\label{eq:hamiltonien}
H&:=&\sum _{j=1}^N\Bigl(-\Delta _{x_j}\, -\, \sum _{k=1}^LZ_k|x_j-R_k|^{-1}\Bigr)\, +\, \sum _{1\leq j<j'\leq N}|x_j-x_{j'}|^{-1}\, +\, E_0\comma\hspace{.4cm}\\
\mbox{where}\ E_0&:=&\sum _{1\leq k<k'\leq L}Z_kZ_{k'}|R_k-R_{k'}|^{-1}\nonumber
\end{eqnarray}
and $-\Delta _{x_j}$ stands for the Laplacian in the variable $x_j$. Here we denote by $|\cdot|$ the euclidian norm on $\R^3$. Setting $\Delta :=\sum _{j=1}^N\Delta _{x_j}$, we define the potential $V$ of the system as the multiplication operator satisfying $H=-\Delta +V$. It is well-known that the Hamiltonian $H$ is a self-adjoint operator on the Sobolev space $\rW ^{2,2}(\R^{3N})$. Let us now fix an electronic bound state $\psi\in \rW ^{2,2}(\R^{3N})\setminus\{0\}$ such that, for some real $E$, $H\psi =E\psi$ (there does exist such a state, see \cite{fh,si,z}). \\
Associated to that bound state $\psi$, we consider the following notions of electronic density (see \cite{e,le,lise,lsc}). 
Let $k$ be an integer such that $0<k<N$. Let $\rho _k : (\R^3)^k\to\R$ be the almost everywhere defined, $\rL^1(\R^{3k})$-function given by, for $\ux=(x_1; \cdots ; x_k)\in\R^{3k}$, 
\begin{equation}\label{eq:rho-densité}
 \rho _k(\ux)\ =\ \int_{\R^{3(N-k)}}\bigl|\psi (\ux; y)\bigr|^2\, dy\period
\end{equation}
It is called the {\em $k$-particle reduced density}. \\
Define also $\gamma _k : (\R^3)^{2k}\to\C$ as the almost everywhere defined, complex-valued function given by, for $\ux=(x_1; \cdots ; x_k)\in\R^{3k}$ and $\ux'=(x_1'; \cdots ; x_k')\in\R^{3k}$, 
\begin{equation}\label{eq:gamma-densité}
 \gamma _k(\ux; \ux')\ =\ \int_{\R^{3(N-k)}}\psi (\ux; y)\, \overline{\psi (\ux'; y)}\, dy\period
\end{equation}
It is called the {\em $k$-particle reduced density matrix}. \\
Thanks to Kato's important contribution in \cite{k}, we know that the bound state $\psi$ is in fact a continuous function. Therefore, the above densities $\rho _k$ and $\gamma _k$ are actually everywhere defined and continuous, and satisfy $\rho _k(\ux)=\gamma _k(\ux; \ux)$, everywhere. \\
We need to introduce the following subsets of $\R^{3k}$. Denoting for a positive integer $p$ by $\ncg 1; p\ncd$ the set of the integers $j$ satisfying $1\leq j\leq p$, the closed set 
\begin{equation}\label{eq:coll-elec-k}
 \cC _k\ :=\ \bigl\{\ux=(x_1; \cdots ; x_k)\in\R^{3k}\, ;\, \exists (j; j')\in\ncg 1; k\ncd ^2\, ;\, j\neq j'\ \mbox{and}\ x_j=x_{j'}\bigr\}
\end{equation}
gathers all possible collisions between the first $k$ electrons. We describe such collisions as ``internal electronic collisions''.  
The closed set 
\begin{equation}\label{eq:coll-nucl-elec-k}
 \cR _k\ :=\ \bigl\{\ux=(x_1; \cdots ; x_k)\in\R^{3k}\, ;\, \exists j\in\ncg 1; k\ncd \, ,\, \exists \ell\in\ncg 1; L\ncd \, ;\, x_j=R_\ell\bigr\}
\end{equation}
groups together all possible collisions of these $k$ electrons with the nuclei. We set 
\begin{equation}\label{eq:sans-coll-k}
 \cU^{(1)}_k\ :=\ \R^{3k}\setminus \bigl(\cC_k\cup\cR_k\bigr)\comma
\end{equation}
which is an open subset of $\R^{3k}$. \\
The set of all possible collisions between particles is then $\cC_N\cup\cR_N$ and the potential $V$ is real analytic precisely on $\R^{3N}\setminus (\cC_N\cup\cR_N)$. Classical elliptic regularity applied to the equation $H\psi=E\psi$ shows that $\psi$ is also real analytic on $\R^{3N}\setminus (\cC_N\cup\cR_N)$ (cf. \cite{h1}). A better regularity for $\psi$ is not expected (and false in some cases), therefore such a regularity for $\rho _k$ and $\gamma_k$ is not clear. It is however granted on some appropriate region. \\
We also need to consider two sets of positions for the first $k$ electrons and introduce the set of all possible collisions between positions of differents sets, namely  
\begin{equation}\label{eq:collisions-extérieures}
 \cC _k^{(2)}\ :=\ \left\{
\begin{array}{l}
 (\ux ; \ux ')\in(\R^{3k})^2\, ;\ \ux=(x_1; \cdots ; x_k)\, ,\ \ux '=(x_1'; \cdots ; x_k')\, ,\, \\
 \\
 \exists (j; j')\in\ncg 1; k\ncd ^2\, ;\  x_j=x_{j'}'
\end{array}
\right\}\period
\end{equation}
A point $(\ux ; \ux ')\in (\R^{3k})^2$ with $k\geq 2$ such that $x_1=x_2=x_1'$ does belong to $\cC _k^{(2)}$. Such a point ``contains'' an internal electronic collision, namely $x_1=x_2$. For any $0<k<N$, we say that a point $(\ux ; \ux ')\in\cC _k^{(2)}$ represents an (several) ``external electronic collision(s)'' if no internal electronic collision occur in $\ux$ nor in $\ux '$. Therefore, the set of so called ``external electronic collisions'' is given by $(\cU^{(1)}_k\times\cU^{(1)}_k\bigr)\cap\cC _k^{(2)}$. \\
We introduce the open subset of $(\R^{3k})^2$ defined by 
\[\cU^{(2)}_k\ :=\ \bigl(\cU^{(1)}_k\times\cU^{(1)}_k\bigr)\setminus\cC _k^{(2)}\period\]
The above mentioned, known regularity results may be summed up in the following way: for any $0<k<N$, the $k$-particle reduced density $\rho _k$ is real analytic on $\cU^{(1)}_k$ and the $k$-particle reduced density matrix $\gamma _k$ is real analytic on $\cU^{(2)}_k=(\cU^{(1)}_k\times\cU^{(1)}_k)\setminus\cC _k^{(2)}$ (see \cite{fhhs1,fhhs2,hs1,j1,j2}). Note that, for each $k$, the smoothness of $\rho_k$ implies the smoothness of the map $\cU^{(1)}_k\ni\ux\donne \gamma _k(\ux; \ux)$. 

Now, we focus on the matrix $\gamma _k$ for $k=N-1$ and want to determine its regularity near a point 
\[(\hat{\ux} ; \hat{\ux} ')\in \bigl(\cU^{(1)}_{N-1}\times\cU^{(1)}_{N-1}\bigr)\cap\cC _{N-1}^{(2)}\comma\]
that represents an (several) external electronic collision(s). What do we precisely mean by 'regularity'? We use the usual class of (integer) regularity $\rC^p$ with $p\in\N$ (see Section~\ref{s:basique}). We do not consider Hölder spaces. For a positive integer $d$, we define the regularity of a continuous map $f : \R^d\dans\C$ near a point $x_0\in\R^d$ by the integer $p$ if, on some neighbourhood $U$ of $x_0$, the function is in the class $\rC^p$ but, for all neighbourhood $V$ of $x_0$ such that $V\subset U$, $f$ does not belong to the class $\rC^{p+1}$ on $V$. \\
Let us take a point 
\[(\hat{\ux}; \hat{\ux}')\ =\ (\hat{x}_1; \cdots ; \hat{x}_{N-1};\, \hat{x}_1'; \cdots ; \hat{x}_{N-1}')\in \bigl(\cU^{(1)}_{N-1}\times\cU^{(1)}_{N-1}\bigr)\cap\cC _{N-1}^{(2)}\period\]
Our study crucially relies on a special decomposition of the bound state $\psi$ near a two-particle collision that was obtained in \cite{fhhs5}. We use such a decomposition near a nucleus-electron collision as well as near an electron-electron collision. This allows us to split the matrix $\gamma _{N-1}$, on a vicinity $V$ of the point $(\hat{\ux}; \hat{\ux}')$, up to some additive smooth contribution, into an appropriate, finite sum of integrals on regions $\cY$ of $\R^3$ such that, on a neighbourhood of $\hat{\ux}$ times $\cY$ and on a neighbourhood of $\hat{\ux}'$ times $\cY$, the bound state $\psi$ may be decomposed as in \cite{fhhs5}. Let us describe this result, that was obtained in Proposition 3.6 in \cite{jn}, in more precise terms (see Proposition~\ref{prop:mod-C-infini} below for details). \\
The set of collisions of $(\hat{\ux}; \hat{\ux}')$ is given by 
\[\cG\ :=\ \bigl\{(j; j')\in\ncg 1;\, N-1\ncd^2\, ;\ \hat{x}_j\, =\, \hat{x}_{j'}'\bigr\}\period\]
Each collision is an external one since $\hat{\ux}\in\cU^{(1)}_{N-1}$ and $\hat{\ux}'\in\cU^{(1)}_{N-1}$. Let $\cD$ be the subset of 
$\ncg 1;\, N-1\ncd$ built of those $j$ such that there exists $j'$ with $(j; j')\in\cG$. For $j\in\cD$, there is a unique integer $k\in\ncg 1;\, N-1\ncd$ such that $\hat{x}_j=\hat{x}_k'$ and this $k$ is denoted by $c(j)$. This defines a map $c : \cD\dans\ncg 1;\, N-1\ncd$ and 
\[\cG\ :=\ \bigl\{(j; c(j))\in\ncg 1;\, N-1\ncd^2\, ;\ j\in\cD\bigr\}\period\]
For $(\ux; \ux')=(x_1; \cdots ; x_{N-1};\, x_1'; \cdots ; x_{N-1}')$ in some vicinity $V$ of $(\hat{\ux}; \hat{\ux}')$, we can write 
\begin{equation}\label{eq:décomp-gamma-coll-ext}
 \gamma _{N-1}(\ux; \ux')\ =\ \sum _{j\in\cD}\, \gamma _{N-1}^{(j)}(\ux; \ux')\, +\, s(\ux; \ux')\comma
\end{equation}
where $s$ is some smooth function and where, for $j\in\cD$, 
\begin{equation}\label{eq:intro-gamma-j}
\gamma _{N-1}^{(j)}(\ux; \ux')\ :=\ \int_{\cY_j}\, \psi (\ux; y)\, \overline{\psi (\ux' ; y)}\, dy
\end{equation}
over some bounded region $\cY_j$ of $\R^3$. For each $j\in\cD$, on $\cY_j$, there is an electron-electron collision at $y=\hat{x}_j$ for the first state $\psi$ in \eqref{eq:intro-gamma-j}, there is an electron-electron collision at $y=\hat{x}_{c(j)}'=\hat{x}_j$ for the second state $\psi$ in \eqref{eq:intro-gamma-j}, and both states $\psi$ can be decomposed as in \cite{fhhs5}. \\
Since the special state decomposition in \cite{fhhs5} is known at two-particle collisions only, one has to ensure that the two copies of the state $\psi$ in the integral defining $\gamma _k$ (cf. \eqref{eq:gamma-densité}) only ``see'' two-particle collisions. This forces $k=N-1$ and requires that $\hat{\ux}$ and $\hat{\ux}'$ do not contain internal electronic collision (see Remark 3.7 in \cite{jn} for details). \\
In \cite{jn} (Section 3), supplementary information on the special state decomposition was given. This information is of great importance for the study of the regularity in \cite{jn} and also here. For the collision between the variable $x_j$ (resp. $x_{c(j)}'$) and the last variable $x_N$ at $\hat{x}_j$ (resp. $\hat{x}_{c(j)}'$), it is described by some nonnegative integer $n_j$ (resp. $n_{c(j)}'$), that we call the {\em relevant valuation} of this collision (see Definition~\ref{def:valuation} and just before Lemma~\ref{lm:change-var} for details). \\
Using the Fourier transform, we shall be able, for each $j\in\cD$, to determine the regularity of $\gamma _{N-1}^{(j)}$ near $(\hat{\ux}; \hat{\ux}')$ in terms of $n_j$ and $n_{c(j)}'$. Then, we obtain the regularity of $\gamma _{N-1}$ from \eqref{eq:décomp-gamma-coll-ext}. The Fourier transform was already the main tool in the analysis in \cite{jn}. Here we actually refine this analysis by an appropriate use of the inverse Fourier transform. This leads to 

\begin{theorem}\label{th:régu-matice-densité}
 Consider a bound state $\psi$ of the $N$-electron, molecular Hamiltonian \eqref{eq:hamiltonien} {\rm (}with $N>1${\rm )} and the associated $(N-1)$-particle reduced density matrix $\gamma _{N-1}$. 
 Let $(\hat{\ux}; \hat{\ux}')\in\bigl(\cU^{(1)}_{N-1}\times\cU^{(1)}_{N-1}\bigr)\cap\cC _{N-1}^{(2)}$. Let 
 \begin{equation}\label{eq:p}
  p\ =\ \min\, \bigl\{n_j+n_{c(j)}';\, j\in\cD\bigr\}\period
 \end{equation}
 There exist a neighbourhood $\cN$ of $\hat{\ux}$, a neighbourhood $\cN'$ of $\hat{\ux}'$, a function $S : \cN\times\cN'\dans\C$ that belongs to $\rC^{5+p}$, and, for $j\in\cD$, for all $\alpha\in\N^3$ with $|\alpha |=n_j$, for all $\alpha'\in\N^3$ with $|\alpha '|=n_{c(j)}'$, real analytic functions $\varphi _{\alpha ; j}$, defined near $\hat{\ux}$, real analytic functions $\varphi _{\alpha '; c(j)}'$, defined near $\hat{\ux}'$, such that, for $(\ux; \ux')\in \cN\times\cN'$, 
\begin{equation}\label{eq:dév-gamma-N-1}
 \gamma _{N-1}(\ux; \ux')\ =\ \sum_{j\in\cD}\, T_j(\ux; \ux')\, +\, S(\ux; \ux')\comma
\end{equation}
where, writing $\ux=(x_j; \ux_j)$ and $\ux '=(x_{c(j)}'; \ux_{c(j)}')$,
\begin{align}
&\ T_j(\ux; \ux')\label{eq:formule-T-j}\\
\ =&\ \frac{-16\pi }{\bigl(6+2n_j+2n_{c(j)}'\bigr)!}\; \sum _{|\alpha |\, =\, n_j\atop|\alpha '|\, =\, n_{c(j)}'}\, 
\varphi_{\alpha ;j}\bigl((x_j+x_{c(j)}')/2; \ux_j\bigr)\, \overline{\varphi '}_{\alpha '; c(j)}\bigl((x_j+x_{c(j)}')/2; \ux_{c(j)}'\bigr)\nonumber\\
&\hspace{5.5cm}\times\, \Bigl(P_\alpha (-\partial_x)\, P_{\alpha '} (\partial_x)\, |x|^{5+2n_j+2n_k'}\Bigr)_{|x=x_j-x_{c(j)}'}\comma\nonumber
\end{align}
where, for $\beta\in\N^3$, the functions $P_\beta$ are some universal homogeneous polynomials of degree $|\beta|$. In particular, the regularity of each $T_j$ near $(\hat{\ux}; \hat{\ux}')$ is precisely $4+n_j+n_{c(j)}'$ and the regularity of $\gamma _{N-1}$ near $(\hat{\ux}; \hat{\ux}')$ is exactly $4+p$. \\
Moreover, for all $j\in\cD$, the real analytic functions $\varphi _{\alpha ; j}$ can be extracted from the special decomposition of the state $\psi$ near the two-particle collision at $(\hat{\ux}; \hat{x}_j)$ and the real analytic functions $\varphi _{\alpha ; c(j)}'$  can be extracted from the special decomposition of the state $\psi$ near the two-particle collision at $(\hat{\ux}'; \hat{x}_{c(j)}')$ with $\hat{x}_{c(j)}'=\hat{x}_j$. 
\end{theorem}
\begin{remark}\label{r:résultat-principal} Several comments on this result can be made. 
\begin{enumerate}
 \item The functions $\varphi _{\alpha ; j}$ {\rm (}resp. $\varphi _{\alpha '; c(j)}'${\rm )} come from the special state decomposition of \cite{fhhs5} for the electron-electron collision of the variables $x_j$ {\rm (}resp. $x_{c(j)}'${\rm )} and $x_N$ at $\hat{x}_j$. See \cite{fhhs5}, Proposition~\ref{prop:décomp-anal}, Remark~\ref{r:décomp-anal-nouv-var}, and the formulae \eqref{eq:dév-varphi-j} and \eqref{eq:dév-varphi'-k}. 
 \item The family $P_\beta$ for $\beta\in\N^3$ is related to the iterated derivatives of the function $\R^3\setminus\{0\}\ni\xi\donne |\xi|^{-4}$. See \eqref{eq:dérivée-alpha-norme-4} for details.
 \item Each term $T_j$ is actually smooth w.r.t. the variables $\ux_j$ and $\ux_{c(j)}'$. The limitation of its regularity comes from the last factor and takes place on the collision set $\{(\ux ; \ux')\in\cN\times\cN'; x_j=x_{c(j)}'\}$. Outside this set, $T_j$ is smooth w.r.t. all variables. To find the regularity of $T_j$, we make use Lemma 3.4 in {\rm \cite{jn}} {\rm (}see Lemma~\ref{lm:régu-valuation} in the present text{\rm )}. 
 \item The term $T_j$ describes the behaviour of the $\gamma _{N-1}^{(j)}$ {\rm (}cf. \eqref{eq:intro-gamma-j}{\rm )} that ``contains'' a collision of the variables $x_j$ and $x_N$ for the first copy of $\psi$ and a collision of variables $x_{c(j)}'$ and $x_N$ for the second copy. While the regularity of $T_j$ is $4+n_j+n_{c(j)}'$, it was shown in {\rm \cite{jn}} {\rm (}cf. Proposition 3.3{\rm )} that the regularity of the first copy {\rm (}resp. the second copy{\rm )} near the collision is $n_j$ {\rm (}resp. $n_{c(j)}'${\rm )}. 
 \item The lack of smoothness of the density matrix $\gamma_{N-1}$ in $\cN\times\cN'$ takes place on 
\[\bigl\{(\ux ; \ux')\in\cN\times\cN';\, \exists\, j\in\cD\, ;\ x_j=x_{c(j)}'\bigr\}\, \subset\, \cC _{N-1}^{(2)}\period\]
 \item Lower and upper bounds on the regularity of the density matrix $\gamma_{N-1}$ was provided in {\rm \cite{jn}} {\rm (}cf. Proposition 5.1 and the proof of Theorem 1.2{\rm )}, only in the case $\hat{\ux}=\hat{\ux}'$.
 \item If $p=0$ then the density matrix $\gamma_{N-1}$ has a ``fifth order cusp'' at $(\hat{\ux}; \hat{\ux}')$. Indeed, the lack of smoothness of the density matrix $\gamma_{N-1}$ there is due to terms containing a factor $|x_j-x_{c(j)}'|^5$. In the case $N=2$, independently of the value of $p$, our result is consistent 
 with the result obtained in \cite{he} and improves it. 
 \item If we consider a bosonic {\rm (}resp. fermionic{\rm)} bounded state $\psi$, it turns out that all $n_j$ and $n_{c(j)}'$ are even {\rm (}resp. odd{\rm )} by Proposition~\ref{prop:val-boson-fermion}. 
 \item In the fermionic case, the density matrix $\gamma_{N-1}$ belongs to the class $\rC^6$ near $(\hat{\ux}; \hat{\ux}')$. 
 In the comparison with the results in {\rm\cite{c1,c2}} when $N=2$, one should be careful, since we discard spin here. 
 \item We point out that our proof of Theorem~\ref{th:régu-matice-densité} relies on rather elementary arguments. It actually gives a more precise result. For any integer $m>4+p$, there exists some function $S_m$ and, for all $j\in\cD$, ``explicit'' functions $T_{j; m}$ such that \eqref{eq:dév-gamma-N-1} holds true with $S$ replaced by $S_m$ and each $T_j$ by $T_{j; m}$. Each $T_{j; m}$ is related to the special state decompositions mentioned in 1. See \eqref{eq:dev-gamma-2}.  
\end{enumerate}
\end{remark}

As a consequence of Theorem~\ref{th:régu-matice-densité}, we now give a result on the operator $\Gamma$ with kernel $\gamma _{N-1}$, viewed as bounded operator from $\rL^2(\R^{3(N-1)})$ to $\rL^2(\R^{3(N-1)})$. Since Theorem~\ref{th:régu-matice-densité} provides local information near a point $(\hat{\ux}; \hat{\ux}')\in\bigl(\cU^{(1)}_{N-1}\times\cU^{(1)}_{N-1}\bigr)\cap\cC _{N-1}^{(2)}$, our second result concerns a ``localisation'' of $\Gamma$. For appropriate cut-off functions $\chi$ and $\chi'$, that localise near $\hat{\ux}$ and $\hat{\ux}'$ respectively, we consider $\chi\Gamma\chi'$, that is the composition of the multiplication by $\chi'$, the action of $\Gamma$, and the multiplication by $\chi$. Our second result states that there exists a unitary transformation $\rU$ on $\rL^2(\R^{3(N-1)})$ such that $\chi\Gamma\chi'\rU$ is a pseudodifferential operator, the symbol of which belongs to a classical class of smooth symbols (see \cite{h3}, p. 65). This was shown in \cite{jn} (cf. Proposition 4.10 and Section 5.2) only in the case $\hat{\ux}=\hat{\ux}'$ ($\rU$ being the identity) and the present result is more precise, as far as the symbol class is concerned.
Thanks to the structure of $U$, we can even show that $\chi\Gamma\chi'$ acts on ``bosons'' (resp. ``fermions'') as a pseudodifferential operator with smooth symbol. We refer to Section~\ref{s:pseudo}, Proposition~\ref{prop:symbole-weyl}, and Remark~\ref{r:dev-asympt-symboles} for details. 

As already pointed out, our results rest on the special decomposition of the bound state $\psi$ at two-particle collisions, that was derived in \cite{fhhs5}. If one can extend this decomposition to other collisions, preserving its ``analytic'' structure, one can reasonably hope to use the arguments of the present paper to get a precise information on the regularity of all the densities $\rho _k$ and all the density matrices $\gamma _k$ at collisions. 

The paper is organized as follows: In Section~\ref{s:basique}, we introduce some notation and recall well-known facts on electronic bound states. In Section~\ref{s:bilateral}, we recall known results on two-particle collisions, in particular the special decomposition from \cite{fhhs5}, and focus on two-electron collisions. We also recall the decomposition \eqref{eq:décomp-gamma-coll-ext} for the matrix $\gamma _{N-1}$, that was obtained in \cite{jn}. Section~\ref{s:Fourier-matrice} is devoted to the Fourier analysis of appropriate localisations of $\gamma _{N-1}$ leading to our proof of Theorem~\ref{th:régu-matice-densité}. In Section~\ref{s:pseudo}, we extract from appropriate localisations of $\Gamma$ a smooth pseudodifferential
structure. We provide technical results and proofs in an 
Appendix at the end of the paper. 

{\bf Acknowledgments:} The author warmly thanks Sébastien Breteaux, Jérémy Faupin, and Victor Nistor, for fruitful discussions and advice.

\section{Notation and well-known facts.}
\label{s:basique}
\setcounter{equation}{0}

We start with a general notation. We denote by $\R$ the field of real numbers and by $\C$ the field of complex numbers.  \\
Let $d$ be a positive integer. For $u\in\R^d$, we write $|u|$ for the euclidian norm of $u$ and we denote by ``$\cdot$'' the corresponding scalar product. Given such a vector $u\in\R^d$ and a nonnegative real number $r$, we denote by $B(u; r[$ (resp. $B(u; r]$) the open (resp. closed) ball of radius $r$ and centre $u$, for the euclidian norm $|\cdot|$ in $\R^d$. \\
In the one dimensional case, we use the following convention for (possibly empty) intervals: for $(a; b)\in\R^2$, let $[a; b]=\{t\in\R; a\leq t\leq b\}$, $[a; b[=\{t\in\R; a\leq t<b\}$, $]a; b]=\{t\in\R; a<t\leq b\}$, and $]a; b[=\{t\in\R; a<t<b\}$.\\
We denote by $\N$ the set of nonnegative integers and set $\N^\ast=\N\setminus\{0\}$. 
If $p\leq q$ are non negative integers, we set $\ncg p; q\ncd :=[p; q]\cap\N$, $\ncg p; q\ncg =[p; q[\cap\N$, $\ncd p; q\ncg =]p; q[\cap\N$, and $\ncd p; q\ncd :=]p; q]\cap\N$.\\
Given an open subset $O$ of $\R^d$ and $n\in\N$, we denote by $W^{n,2}(O)$ the standard Sobolev space of those $\rL^2$-functions on $O$ such that, for $n'\in \ncg 0; n\ncd$, their distributional partial derivatives of order $n'$ belong to $\rL^2(O)$. In particular, $W^{0,2}(O)=\rL^2(O)$. Without reference to $O$, we denote by $\|\cdot\|$ (resp. $\langle\cdot , \cdot\rangle$) the $\rL^2$-norm (resp. the right linear scalar product) on $\rL^2(O)$. \\
On $\R^d$, we use a standard notation for partial derivatives. For $j\in\ncg1; d\ncd$, we denote by $\partial _j$ or $\partial _{\rx_j}$ the $j$'th first partial derivative operator. For $\alpha\in\N^d$ and $\rx\in\R^d$, we set $D_\rx^\alpha :=(-i\partial _\rx)^\alpha :=(-i\partial _{\rx_1})^{\alpha _1}\cdots(-i\partial _{\rx_d})^{\alpha _d}$, $D_\rx=-i\nabla_\rx$, $\rx^\alpha :=\rx_1^{\alpha_1}\cdots \rx_d^{\alpha_d}$, $|\alpha |:=\alpha_1+\cdots +\alpha _d$, $\alpha !:=(\alpha_1!)\cdots (\alpha_d !)$, $|\rx|^2=\rx_1^2+\cdots +\rx_d^2$, and $\langle \rx\rangle :=(1+|\rx|^2)^{1/2}$. Given $(\alpha ; \beta)\in(\N^d)^2$, we write $\alpha\leq\beta$ if, for all $j\in\ncg 1; d\ncd$, $\alpha _j\leq\beta_j$. In that case, we define the multiindex $\beta -\alpha:=(\beta_j-\alpha _j)_{j\in\ncg 1; d\ncd}\in \N^d$. \\
We choose the same notation for the length $|\alpha|$ of a multiindex $\alpha\in\N^d$ and for the euclidian norm $|\rx|$ of a vector $\rx\in\R^d$ but the context should avoid any confusion. \\
For $k\in\N\cup\{\infty\}$, we denote by $\rC^k(O)$ the vector space of functions from $O$ to $\C$ which have continuous derivatives up to order $k$ and by $\rC_c^k(O)$ the intersection of $\rC^k(O)$ with the set of functions with compact support in $O$. 
If a function $f$ satisfies $f\in\rC^k(O)$ with $k\in\N\cup\{\infty\}$, we often write for this that the function $f$ belongs to the class $\rC^k$ on $O$. In the case $k=\infty$, we also write that $f$ is smooth on $O$ if $f\in\rC^\infty(O)$. \\
Let $\rx_0\in O$. For all $\alpha\in\N^{d}$, let $a_\alpha\in\C$ and $f_{\rx_0;\, \alpha} : O\dans\C$ be defined by $f_{\rx_0;\, \alpha} (\rx)=a_\alpha(\rx-\rx_0)^\alpha$. The associated power series $\sum_{\alpha\in\N^{d}}f_{\rx_0;\, \alpha}$ is the sequence of functions on $O$ of the form 
\[\left(\sum_{\alpha\in\N^{d},\, |\alpha|\leq N}\, f_{\rx_0;\, \alpha}\right)_{N\in\N}\period\]
Let $U$ be the set of the $\rx\in\R^d$ such that the previous sequence at $\rx$ converges in $\C$. The set $U$ contains at least $\rx_0$. The sum of this power series is the map $\varphi : U\dans\C$ defined 
\[\forall\, \rx\in U\comma\hspace{.4cm}\varphi (\rx)\ =\ \lim _{N\to\infty}\, \sum_{\alpha\in\N^{d},\, |\alpha|\leq N}\, f_{\rx_0;\, \alpha}(\rx)\ =\ \lim _{N\to\infty}\, \sum_{\alpha\in\N^{d},\, |\alpha|\leq N}\, a_\alpha(\rx-\rx_0)^\alpha\period\]
A function $f: O\to\C$ is real analytic if, for any $\rx_0\in O$, there exists some neighbourhood $U$ of $\rx_0$ such that $f$ coincides on $U$ with the sum of some power series $\sum _{\alpha\in\N^{d}}f_{\rx_0;\, \alpha}$. Real analytic functions on $O$ are smooth on $O$. We say that a real analytic function $f$ is zero if it is the zero map on its domain of definition (i.e. if it is identically zero) and write $f=0$ in this case. If this is not the case, we write $f\neq 0$. We refer to \cite{ca,h4} for details on the (real) analyticity w.r.t. several variables. \\
We shall frequently use standard ``continuity and partial derivation under the integral sign''. See \cite{d,m} for details. \\
Integration by parts in integrals will be used and often combined with identities of the following types: 
\begin{equation}\label{eq:dérivée-exp}
 \forall (x; \xi)\in \R^d\times\bigl(\R^d\setminus\{0\}\bigr)\comma\hspace{.4cm}-i\, \frac{\xi}{|\xi|^2}\, \cdot\, \nabla_{x}\, e^{i\, \xi\cdot x}\ =\ e^{i\, \xi\cdot x}
\end{equation}
and 
\begin{equation}\label{eq:dérivée-exp-crochet}
 \forall (x; \xi)\in \R^d\times\R^d\comma\hspace{.4cm}\langle\xi\rangle^{-2}\bigl(1\, +\, (1/i)\, \xi\cdot \nabla _x\bigr)\, e^{i\xi\cdot x}\ =\ e^{i\xi\cdot x}\period
\end{equation}
We shall make use of the usual Fourier transform on $\R^d$. Given an integrable function $g : \R^d\dans\C$, its Fourier transform is the continuous, bounded map $F_g : \R^d\dans\C$ defined by
\begin{equation}\label{eq:fourier-g}
 \forall\, \xi\in\R^d\comma\hspace{.4cm}F_g(\xi)\ =\ \int_{\R^d}\, e^{-i\, \xi\cdot x}\; g(x)\, dx
\end{equation}
while its inverse Fourier transform is the continuous, bounded map $F_g^i : \R^d\dans\C$ defined by
\begin{equation}\label{eq:fourier-inv-g}
 \forall\, \xi\in\R^d\comma\hspace{.4cm}F_g^i(\xi)\ =\ (2\pi)^{-d}\, \int_{\R^d}\, e^{i\, \xi\cdot x}\; g(x)\, dx\period
\end{equation}
If $F_g$ is integrable then $g$ is continuous and is given by the inverse Fourier transform of $F_g$, that is 
\begin{equation}\label{eq:fourier-inv-g-bis}
 \forall\, x\in\R^d\comma\hspace{.4cm}g(x)\ =\ (2\pi)^{-d}\, \int_{\R^d}\, e^{i\, \xi\cdot x}\; F_g(\xi)\, d\xi\period
\end{equation}
We shall use the following elementary lemma, that is a slightly modified version of Lemma 4.4 in \cite{jn}. 
\begin{lemma}\label{lm:inté-parties}
Let $d\in\N^\ast$ and $k\in\N$. Let $g: \R^d\dans\C$ be an integrable function. We denote by $F_g$ its Fourier transform. Given a real $r$, we denote by $E(r)$ the integer part of $r$, that is the biggest integer less or equal to $r$. 
\begin{enumerate}
 \item If the function $g: \R^d\dans\C$ belongs to the class $\rC^k$ and is compactly supported then $F_g$ is smooth and there exists $C>0$ such that, for all $\xi\in\R^d$ with $|\xi|\geq 1$, 
 \[\bigl|F_g(\xi)\bigr|\ \leq\ C\; |\xi|^{-k}\period\]
 \item Assume that the function $F_g: \R^d\dans\C$ belongs to the class $\rC^0$ and satisfies, for some real $r>E(r)\geq 0$ and some $C>0$, for all $\xi\in\R^d$ with $|\xi|\geq 1$, 
 \[\bigl|F_g(\xi)\bigr|\ \leq\ C\; |\xi|^{-r-d}\period\]
 Then the function $g: \R^d\dans\C$ belongs to the class $\rC^{E(r)}$. 
\end{enumerate}
\end{lemma}
We shall also exploit another elementary, but important lemma, namely 
\begin{lemma}\label{lm:fourier-norme-localisée}{\rm \cite{jn} (}Lemma 4.5{\rm )}.
Let $q\in\N$ and $f : \R^3\ni x\donne |x|^{2q+1}\cdot\tau \bigl(|x|\bigr)$ where $\tau\in\rC_c^\infty(\R)$ such that $\tau =1$ near $0$. Then, its Fourier transform $F_f$ is a real analytic, bounded function on $\R^3$, which is given, for $\xi\neq 0$, by 
\begin{equation}\label{eq:fourier-norme-localisée}
F_f(\xi)\ =\ \frac{4\pi}{|\xi|}\, \int_0^{+\infty}\, \tau (r)\; r^{2q+2}\; \sin (r|\xi|)\, dr\period
\end{equation}
It has the following behaviour at infinity: 
\begin{equation}\label{eq:fourier-norme-localisée-symbole-1}
 \forall\, \alpha\in\N^3\comma\ \exists\, C_\alpha>0\, ;\ \forall\, \xi\in \R^3\setminus\{0\}\comma\ \bigl|\partial^\alpha F_f(\xi)\bigr|\ \leq\ C_\alpha\;  |\xi |^{-4-2q-|\alpha|}\period
\end{equation}
Furthermore, there exists a smooth function $G :\R^3\setminus\{0\}\dans\R$ such that, for $\xi\neq 0$,
\[F_f(\xi )\ =\ 4\pi (-1)^{q+1}\bigl((2q+2)!\bigr)\, |\xi |^{-4-2q}\, +\, G(\xi )\]
and such that, for all $k\in\ncg 4+2q+1; +\infty\ncg$ and $\alpha\in\N^3$, 
\begin{equation}\label{eq:fourier-norme-localisée-symbole-2}
 \exists\, C_{k; \alpha}>0\, ;\ \forall\, \xi\in \R^3\setminus\{0\}\comma\ \bigl|\partial^\alpha G(\xi)\bigr|\ \leq\ C_{k; \alpha}\;  |\xi |^{-k-|\alpha|}\period
\end{equation}
\end{lemma}
\Pf The case $q=0$ was stated and proved in \cite{jn}. The proof there actually extends to the general case. By Theorem 7.1.14 in \cite{h2}, $F_f$ is real analytic. $\cqfd$ 

Thanks to Hardy's inequality 
\begin{equation*}
\exists\, c>0\, ;\ \forall\, f\in \rW^{1,2}(\R^3)\comma \ \int _{\R^3}|t|^{-2}\, |f(t)|^2\, dt\ \leq \ c\int _{\R^3}|\nabla f(t)|^2\, dt\comma 
\end{equation*}
one can show that $V$ is $\Delta$-bounded with relative bound $0$. Therefore the Hamiltonian $H$ is self-adjoint on the domain of the Laplacian $\Delta $, namely $\rW ^{2,2}(\R^{3N})$ (see Kato's theorem in \cite{rs2}, p. 166-167). 
We point out (cf.\ \cite{si,z}) that a bound state $\psi$ exists at least for appropriate $E\leq E_0$ (cf. \cite{fh}) and for 
$N<1+2\sum _{k=1}^LZ_k$. A priori, it belongs to the Sobolev space $\rW ^{2,2}(\R^{3N})$, a space that contains non-continuous functions. But, as shown in $\cite{k}$, $\psi$ is actually continuous. Since the integrand in \eqref{eq:rho-densité} (resp. in \eqref{eq:gamma-densité}) is integrable and continuous, a standard result on the continuity of integrals depending on parameters shows that $\rho _k$ (resp. $\gamma _k$) is everywhere defined and continuous. \\
Finally we recall further, well-know properties of a bound state of $H$ (see Section 2 in \cite{jn} for details). 
\begin{proposition}\label{prop:faits-état}
Recall that $\cC_N$ {\rm (}resp. $\cR_N${\rm )} is defined in \eqref{eq:coll-elec-k} {\rm (}resp. \eqref{eq:coll-nucl-elec-k}{\rm )}. 
The bound state $\psi$ is a continuous function that also belongs to the Sobolev space $\rW ^{2,2}(\R^{3N})$.
On the open set $\R^{3N}\setminus (\cC_N\cup\cR_N)$, $\psi$ is a real analytic function. Take a subset $\cE$ of $\R^{3N}\setminus (\cC_N\cup\cR_N)$ such that its distance to the collisions set $\cC_N\cup\cR_N$ is positive. Then any partial derivative of $\psi$ belongs to $\rL^2(\cE)$. For any non-empty open set $\cO$ of $\R^{3N}$, the bound state $\psi$ does not vanish identically on $\cO$. 
\end{proposition}
%

\section{Two-electron collisions.}
\label{s:bilateral}
\setcounter{equation}{0}

In this section, we recall several results, that were obtained in \cite{jn} and are based on the special state decomposition at a two-particle collision from \cite{fhhs5}. We first focus on two-particle collisions of electrons and introduce the notion of relevant valuation associated to the collision. Then we present the decomposition \eqref{eq:décomp-gamma-coll-ext} of $\gamma _{N-1}$ in details. We use basic notions of real analytic functions of several variables (see \cite{h4}). 

\begin{proposition}\label{prop:décomp-anal}{\rm \cite{jn} (}Proposition 3.3{\rm )}.\\
The set $\cR_N$ being defined in \eqref{eq:coll-nucl-elec-k}, we consider a point $\hat{\uz}=(\hat{z}_1;\, \cdots\, ;\, \hat{z}_N)\in\R^{3N}\setminus\cR_N$ such that there exists an unique $(j; k)\in\ncg 1;\, N\ncd^2$ such that $\hat{z}_j=\hat{z}_k$ and $j\neq k$. 
According to \eqref{eq:coll-elec-k}, $\hat{\uz}\in\cC_N$ and a two-electron collision occurs at $\hat{\uz}$. 
Then there exists a neighbourhood $\Omega$ of $\hat{\uz}$ in $\R^{3N}$ and two sums of power series $\tilde\varphi _1$ and $\tilde\varphi _2$ on $\Omega$ such that, setting $\uz=(z_1;\, \cdots\, ;\, z_N)\in\R^{3N}$, 
\begin{equation}\label{eq:décomp-électrons}
 \forall \uz\in\Omega\comma\hspace{.4cm}\psi (\uz)\ =\ \tilde\varphi _1(\uz)\, +\, (1/2)\, |z_j\, -\, z_k|\; \tilde\varphi _2(\uz)\period
\end{equation}
Furthermore, the function $\tilde\varphi _2$ is not zero. Both functions $\tilde\varphi _1$ and $\tilde\varphi _2$ are uniquely determined by $\psi$ and the two-electron collision $\hat{\uz}$.
\end{proposition}
In this situation, the nonzero function $\tilde\varphi _2$ may vanish on the collision set $\{\uz\in\Omega ;\, z_j=z_k\}$. If it does, it will be important to describe how it vanishes. For instance, is it like $|z_j-z_k|$ or like $|z_j-z_k|^3$? To this end, we use the notion of {\em relevant valuation} at the two-electron collision $\hat{\uz}$, that was introduced in Definition 3.2 in \cite{jn}. \\
Given a nonzero, real analytic function $\varphi$ in several variables $\uz=(z_1;\, \cdots\, ;\, z_N)\in\R^{3N}$, it may be written, near any point $\hat{\uz}=(\hat{z}_1;\, \cdots\, ;\, \hat{z}_N)$ of its domain of analyticity, as the sum of a power series in the variables $((z_1-\hat{z}_1);\, \cdots\, ;\, (z_N-\hat{z}_N))$. For $j\in\ncg 1;\, N\ncd$, this sum may be rearranged in the following form 
\begin{equation}\label{eq:série-ent-j}
\varphi (\uz )\ =\ \sum_{\alpha\in\N^3}\, \varphi _{\alpha}\bigl((z_k)_{k\neq j}\bigr)\; (z_j\, -\, \hat{z}_j)^{\alpha }\comma 
\end{equation}
for sums $\varphi _{\alpha}$ of appropriate power series in the variables $z_k$ with $k\neq j$. Since the function $\varphi$ is nonzero, so is at least one function $\varphi _{\alpha}$. This means that the set $\{|\alpha|;\, \alpha\in\N^3,\, \varphi _\alpha\neq 0\}$ is a non empty subset of $\N$. By definition, the valuation of $\varphi$ in the variable $z_j$ at $\hat{\uz}$ is the minimum of this set. When $\varphi$ is zero, we decide to set its valuation in the variable $z_j$ at $\hat{\uz}$ to $-\infty$. 

\begin{definition}\label{def:valuation}
Let $\hat{\uz}=(\hat{z}_1;\, \cdots\, ;\, \hat{z}_N)\in\cC_N\setminus\cR_N$ as in Proposition~\ref{prop:décomp-anal}. We introduce new variables by setting, on $\Omega$, $\mfz_\ell =z_\ell$, if $\ell\not\in\{j; k\}$, $\mfz_j=(z_j-z_k)/2$, and $\mfz_k=(z_j+z_k)/2$. Replacing each $z_j$ by $\hat{z}_j$, we similarly define the $\hat{\mfz}_\ell$ from $\hat{z}_\ell$ and set $\hat{\umfz}=(\hat{\mfz}_1;\, \cdots\, ;\, \hat{\mfz}_N)$. The sum of a power series $\tilde\varphi _2$ on $\Omega$ may be rewritten as $\umfz\donne\varphi _2(\umfz)$, for $\umfz=(\umfz_1;\, \cdots\, ;\, \umfz_N)$ in some neighbourhood of $\hat{\umfz}$ and for some sum of a power series $\varphi _2$ near $\hat{\umfz}$. In that case, we define the {\em relevant valuation} of the two-electron collision at $\hat{\uz}$ as the valuation of $\varphi _2$ in the variable $\mfz_j$ at $\hat{\umfz}$. 
\end{definition}
\begin{remark}\label{r:décomp-anal-nouv-var}Under the assumptions of Proposition~\ref{prop:décomp-anal}, we can use the new variables, introduced in Definition~\ref{def:valuation}, to rewrite \eqref{eq:décomp-électrons} as 
\begin{equation}\label{eq:décomp-électrons-bis}
 \forall \uz\in\Omega\comma\hspace{.4cm}\psi (\uz)\ =\ \varphi _1\bigl(\umfz\bigr)\, +\, |\mfz_j|\; \varphi _2\bigl(\umfz\bigr)
\end{equation}
where, if $j<k$,  
\[\psi (\uz)\ =\ \psi \bigl(\mfz_1; \cdots ; \mfz_{j-1};\, \mfz_k+\mfz_j\, ; \mfz_{j+1}; \cdots ; \mfz_{k-1}; \, \mfz_k-\mfz_j\, ; \mfz_{k+1};\cdots ; \mfz_N\bigr)\]
and where $\varphi _1$ and $\varphi _2$ are sums of a power series. 
We observe that, when $\uz$ runs in $\Omega$, the variables $(\mfz_k)_{\ell\not\in\{j; k\}}$ runs in some neighbourhood of $(\hat{z}_k)_{\ell\not\in\{j; k\}}$, while the variable $\mfz_j$ runs in some neighbourhood of $0$ and the variable $\mfz_k$ runs in some neighbourhood of $\hat{z}_j=\hat{z}_k$. \\
For simplicity, we set, for $\uz\in\Omega$ and $j<k$, 
\[\bigl(z_j; z_k;\, Z_{j; k}\bigr)\ :=\ \bigl(z_1; \cdots ; z_{j-1};\, z_j\, ; z_{j+1}; \cdots ; z_{k-1}; \, z_k\, ; z_{k+1};\cdots ; z_N\bigr)\]
where $Z_{j; k}=(z_\ell)_{\ell\not\in\{j; k\}}$. 
\end{remark}

In the framework of Proposition~\ref{prop:décomp-anal}, it was proved in \cite{jn} (cf. Proposition 3.3) that, if $n$ denotes the relevant valuation 
of the two-electron collision, then the regularity (in the above sense) of the bound state $\psi$ near this collision is exactly $n$. 
This was derived from the following result, that we shall use here also. 
\begin{lemma}\label{lm:régu-valuation}{\rm \cite{jn}(}Lemma 3.4{\rm )}.\\ 
Let $\cW$ be a bounded neighbourhood of $0$ in $\R^3$ and $\varphi :\cW\dans\C$ be a nonzero real analytic function with valuation $q\in\N$ {\rm (}w.r.t. its $3$-dimensional variable in the above sense{\rm )}. Then the function $N_\varphi : \cW\ni x\donne |x|\, \varphi (x)\in\C$ belongs to the class $\rC^q$ but does not belong to the class $\rC^{q+1}$. Furthermore, any partial derivative of order $q+1$ of $N_\varphi$ is well-defined away from zero and bounded. 
\end{lemma}

It was claimed without proof in \cite{jn} (cf. Remark 3.5) that, if the state $\psi$ is bosonic (resp. fermionic) then so are the 
functions $\tilde\varphi _1$ and $\tilde\varphi _2$ appearing in Proposition~\ref{prop:décomp-anal} and therefore the relevant valuation of the two-electron collision is even (resp. odd). We now prove this. \\
Recall that the bound state $\psi$ is bosonic if it is invariant under any exchange of two electronic coordinates, that is, for the exchange of $z_j$ and $z_k$,  
\begin{align}
&\psi \bigl(z_1; \cdots ; z_{j-1};\, z_k\, ; z_{j+1}; \cdots ; z_{k-1}; \, z_j\, ; z_{k+1};\cdots ; z_N\bigr)\nonumber\\
=\ &\psi \bigl(z_1; \cdots ; z_{j-1};\, z_j\, ; z_{j+1}; \cdots ; z_{k-1}; \, z_k\, ; z_{k+1};\cdots ; z_N\bigr)\comma\label{eq:boson}
\end{align}
for all $(z_1; \cdots ; z_N)\in (\R^3)^N$. It is fermionic if any exchange of two electronic coordinates changes its sign, that is, for the exchange of $z_j$ and $z_k$,   
\begin{align}
&\psi \bigl(z_1; \cdots ; z_{j-1};\, z_k\, ; z_{j+1}; \cdots ; z_{k-1}; \, z_j\, ; z_{k+1};\cdots ; z_N\bigr)\nonumber\\
=\ -\, &\psi \bigl(z_1; \cdots ; z_{j-1};\, z_j\, ; z_{j+1}; \cdots ; z_{k-1}; \, z_k\, ; z_{k+1};\cdots ; z_N\bigr)\comma\label{eq:fermion}
\end{align}
for all $(z_1; \cdots ; z_N)\in (\R^3)^N$ (cf. \cite{rs1} p. 53-54 or \cite{lise} p. 35). 

\begin{proposition}\label{prop:val-boson-fermion}
In the framework of Proposition~\ref{prop:décomp-anal}, we consider a bound state $\psi$ that is bosonic {\rm (}resp. fermionic{\rm )}. Then, on an appropriate vicinity of $\hat{\uz}$, that is included in $\Omega$, \eqref{eq:boson} {\rm (}resp. \eqref{eq:fermion}{\rm )} holds true with $\psi$ replaced by $\tilde\varphi _1$ and also with $\psi$ replaced by $\tilde\varphi _2$. In particular, the relevant valuation of the two-electron collision at $\hat{\uz}$ is even {\rm (}resp. odd{\rm )}. 
\end{proposition}
\Pf Let $r>0$ and, for $\ell\in\ncg 1; N\ncd$, let $\cV_\ell$ be the ball of radius $r$ and centre $\hat{z}_\ell$, such that the cartesian product $C:=\cV_1\times\cdots\times\cV_N$ is included in $\Omega$. The set $C$ is invariant under the exchange of the coordinates $z_j$ and $z_k$. By assumption, we have, for some $\epsilon\in\{-1; 1\}$, for all $\uz\in C$, 
\begin{equation}\label{eq:sym}
\psi\bigl(z_k; z_j;\, Z_{j; k}\bigr)\ :=\ \epsilon\, \psi\bigl(z_j; z_k;\, Z_{j; k}\bigr)\comma
\end{equation}
where $\epsilon =1$, if $\psi$ is bosonic and, $\epsilon =-1$, if $\psi$ is fermionic. 
By \eqref{eq:décomp-électrons}, we get, for $\uz\in C$, 
\begin{align}
f\bigl(z_k; z_j;\, Z_{j; k}\bigr)\ =&\ \frac{|z_j\, -\, z_k|}{2}\; g\bigl(z_k; z_j;\, Z_{j; k}\bigr)\label{eq:anal-non-anal}\\
\mbox{where}\hspace{.4cm} f\bigl(z_k; z_j;\, Z_{j; k}\bigr)\ :=&\ \tilde\varphi _1\bigl(z_k; z_j;\, Z_{j; k}\bigr)\, -\, \epsilon\, \tilde\varphi _1\bigl(z_j; z_k;\, Z_{j; k}\bigr)\nonumber\\
\mbox{and}\hspace{.4cm} g\bigl(z_k; z_j;\, Z_{j; k}\bigr)\ :=&\ \epsilon\, \tilde\varphi _2\bigl(z_j; z_k;\, Z_{j; k}\bigr)\, -\, \tilde\varphi _2\bigl(z_k; z_j;\, Z_{j; k}\bigr)\period\nonumber
\end{align}
For fixed $z_k$ and fixed $Z_{j;k}$, the map $x\donne f(z_k+x; z_k; Z_{j;k})$ is smooth. By \eqref{eq:anal-non-anal} and Lemma~\ref{lm:régu-valuation}, the real analytic map $x\donne g(z_k+x; z_k; Z_{j;k})$ must be zero. Since this holds true for all 
$z_k$ and all $Z_{j;k}$, $f$ and $g$ are zero. This shows that $\tilde\varphi _1$ and $\tilde\varphi _2$ satisfy \eqref{eq:boson}  (resp. \eqref{eq:fermion}) if $\epsilon =1$ (resp. $\epsilon =-1$). \\
Now we use the change of variables of Remark~\ref{r:décomp-anal-nouv-var} and the symmetry of $\tilde\varphi _2$. 
For $\umfz$ in an appropriate neighbourhood of $(0; \hat{z}_k; \hat{Z}_{j; k})$, 
\[\varphi _2\bigl(-\mfz _j; \mfz _k; Z_{j; k}\bigr)=\epsilon\, \varphi _2\bigl(\mfz _j; \mfz _k; Z_{j; k}\bigr)\period\]
Using the expansion \eqref{eq:série-ent-j} for $\varphi =\varphi _2$, we obtain 
\[0\ =\ \sum _{\alpha\in\N^3}\, \bigl((-1)^{|\alpha|}\, -\, \epsilon\bigr)\, \varphi _{\alpha}\bigl((\mfz_\ell)_{\ell\neq j}\bigr)\; \mfz_j^{\alpha}\period\]
Thus all the $\varphi _{\alpha}$ for odd $|\alpha|$ are zero if $\epsilon =1$ and all the $\varphi _{\alpha}$ for even $|\alpha|$ are zero if $\epsilon =-1$. Therefore the relevant valuation, that is the valuation of $\varphi _2$ in the variable $\mfz_j$ at $\hat{\umfz}$, is even if $\epsilon =1$ and odd if $\epsilon =-1$.
$\cqfd$

Making use of the special state decompositions in \cite{fhhs5} at a nucleus-electron collision and at an electron-electron collision, the decomposition \eqref{eq:décomp-gamma-coll-ext} was derived in Proposition 3.6 in \cite{jn}. We need some notation. \\
Recall that we introduced the set $\cU^{(1)}_{N-1}$ in \eqref{eq:sans-coll-k} and the set $\cC _{N-1}^{(2)}$ in \eqref{eq:collisions-extérieures}. Let us take $(\hat{\ux}; \hat{\ux}')\in (\cU^{(1)}_{N-1}\times\cU^{(1)}_{N-1})\, \cap\, \cC _{N-1}^{(2)}$. As vectors in $\R^{3(N-1)}$, we write $\hat{\ux}=(\hat{x}_1;\, \cdots;\, \hat{x}_{N-1})$ and $\hat{\ux}'=(\hat{x}_1';\, \cdots;\, \hat{x}_{N-1}')$. The set of collisions is 
\[\cG\ =\ \bigl\{(j; j')\in\ncg 1;\, N-1\ncd^2\, ;\ \hat{x}_j\, =\, \hat{x}_{j'}'\bigr\}\period\]
Since $(\hat{\ux}; \hat{\ux}')\in \cC _{N-1}^{(2)}$, $\cG$ is not empty. Since $\hat{\ux}\in\cU^{(1)}_{N-1}$ and $\hat{\ux}'\in\cU^{(1)}_{N-1}$, $\cG$ is the graph of an injective map $c : \cD\dans\ncg 1;\, N-1\ncd$ with the domain of definition 
\[\cD\ :=\ \bigl\{j\in\ncg 1;\, N-1\ncd\, ;\ \exists\, j'\in\ncg 1;\, N-1\ncd\, ;\ (j; j')\in\cG\bigr\}\ \neq\ \emptyset\period\]
Using \cite{fhhs5} and Proposition~\ref{prop:faits-état}, we have 
\begin{proposition}\label{prop:mod-C-infini}{\rm \cite{jn} (}Proposition 3.6{\rm )}.\\
 Let $(\hat{\ux}; \hat{\ux}')\in (\cU^{(1)}_{N-1}\times\cU^{(1)}_{N-1})\, \cap\, \cC _{N-1}^{(2)}$. Then there exist an open  neighbourhood $\cV$ of $\hat{\ux}$, an open neighbourhood $\cV'$ of $\hat{\ux}'$, a smooth function $s: \cV\times\cV'\dans\C$, and, for $j\in\cD$, an open neighbourhood $\cW_j$ of $\hat{x}_j$, a function $\tilde\chi _j\in\rC_c^\infty(\R^3; \R)$, and two sums of power series $\tilde\varphi _j : \cV\times \cW_j\dans\C$ and $\tilde\varphi _{c(j)}' : \cV'\times \cW_j\dans\C$, such that $\tilde\chi _j=1$ near $\hat{x}_j=\hat{x}_{c(j)}'$ and the support of $\tilde\chi _j$ is included in $\cW_j$, and such that, for all $(\ux; \ux')\in (\cV\times\cV')$, \eqref{eq:décomp-gamma-coll-ext} holds true, namely 
 \[
\gamma _{N-1}(\ux; \ux')\ =\ s(\ux; \ux')\, +\, \sum _{j\in\cD}\, \gamma _{N-1}^{(j)}(\ux; \ux')\comma
\]
 where 
 \begin{equation}\label{eq:gamma-j}
 \gamma _{N-1}^{(j)}(\ux ;\, \ux ')\ =\ \int_{\R^3}\, |x_j\, -\, y|\;  \tilde\varphi_j(\ux ; y) \; |x_{c(j)}'\, -\, y|\; \overline{\tilde\varphi '}_{c(j)}(\ux '; y)\; \tilde\chi _j(y)\, dy\comma
 \end{equation}
 for $j\in\cD$. For such $j$, the function $\tilde\varphi _j$ is precisely the function $\tilde\varphi_2$ in formula \eqref{eq:décomp-électrons} near $(\hat{\ux}; \hat{x}_j)$ and the function $\tilde\varphi _{c(j)}'$ is precisely the function $\tilde\varphi_2$ in formula \eqref{eq:décomp-électrons} near $(\hat{\ux}'; \hat{x}_j)$. Furthermore, one may require that, for $(j; \ell)\in\cD^2$ with $j\neq\ell$, $\cW _j\cap\cW_\ell=\emptyset$, and that, for $j\in\cD$, $\cW_j$ is included in the range of the map $\cV\ni\ux\donne x_j$ and also included in the range of the map $\cV'\ni\ux '\donne x_{c(j)}'$. 
\end{proposition}
%

\section{Fourier transform of localisations of the density matrix.}
\label{s:Fourier-matrice}
\setcounter{equation}{0}

In order to find the regularity of $\gamma _{N-1}$ near a point $(\hat{\ux}; \hat{\ux}')\in (\cU^{(1)}_{N-1}\times\cU^{(1)}_{N-1})\, \cap\, \cC _{N-1}^{(2)}$, it suffices to consider $\gamma _{N-1}$ multiplied by a cut-off function that is $1$ near $(\hat{\ux}; \hat{\ux}')$. We shall require that the support of this cut-off function is so small that we can exploit Proposition~\ref{prop:mod-C-infini}. Now we want to study the Fourier transform \eqref{eq:fourier-g} of the localised version of $\gamma _{N-1}$ and, by \eqref{eq:décomp-gamma-coll-ext}, it suffices to do so for each $\gamma _{N-1}^{(j)}$, that is defined in \eqref{eq:gamma-j}. 
Using inverse Fourier transform \eqref{eq:fourier-inv-g}, we shall get an expansion of each $\gamma _{N-1}^{(j)}$, from which we shall be able to derive our main result Theorem~\ref{th:régu-matice-densité}.

Let $(\hat{\ux}; \hat{\ux}')\in (\cU^{(1)}_{N-1}\times\cU^{(1)}_{N-1})\, \cap\, \cC _{N-1}^{(2)}$. Recall that, for $\ux\in(\R^3)^{N-1}$ and $q\in\ncg 1; N-1\ncd$, we sometimes write $\ux=(x_q; \ux _q)$. We use the objects introduced in Proposition~\ref{prop:mod-C-infini}. For $j\in\ncg 1; N-1\ncd$, let $\chi _{0;j}\in\rC_c^\infty (\R^3; \R)$ satisfying the following requirements:
\begin{itemize}
 \item If $j\not\in\cD$, $\chi _{0;j}=1$ near $\hat{x}_j$; 
 \item If $j\in\cD$, $\chi _{0;j}=1$ near $\hat{x}_j$ and $\chi _{0;j}=\chi _{0;j}\tilde\chi _j$; 
 \item For $(j; \ell)\in\ncg 1; N-1\ncd^2$ with $j\neq\ell$, $\chi _{0;j}\chi _{0;\ell}=0$. 
\end{itemize}
Let $\chi _0$ be the tensor product $\otimes_{j\in\ncg 1; N-1\ncd}\, \chi _{0;j}$, that is the map 
\[(\R^3)^{N-1}\ni \ux\ \donne\ \prod _{j\in\ncg 1; N-1\ncd}\, \chi _{0;j}(x_j)\period\]
Note that $\chi _0\in\rC_c^\infty((\R^3)^{N-1}; \R)$. Recall that, for $j\in\cD$, the support of $\tilde\chi _j$ is included in $\cW_j$. Now, we can choose the support of the functions $\chi _{0;j}$, for $j\in\ncg 1; N-1\ncd$, so small such that, for $j\in\cD$, there exists an open neighbourhood $\cV_{\neq j}$ of $\hat{\ux}_j$ such that $\cW_j\times\cV_{\neq j}\subset\cV$ and the support of 
$\otimes_{\ell\in\ncg 1; N-1\ncd\setminus\{j\}}\, \chi _{0;\ell}$ is included in $\cV_{\neq j}$. This ensures, in particular, that the support of $\chi_0$ is included in $\cV$. \\
For $k\in\ncg 1; N-1\ncd$, let $\chi _{0;k}'\in\rC_c^\infty (\R^3; \R)$ satisfying the following requirements:
\begin{itemize}
 \item If $k\not\in c(\cD)$, $\chi _{0;k}'=1$ near $\hat{x}_k'$; 
 \item If $k=c(j)$ for $j\in\cD$, $\chi _{0;k}'=1$ near $\hat{x}_{c(j)}'=\hat{x}_j$ and $\chi _{0;k}'=\chi _{0;k}'\tilde\chi _j$. 
 \item For $(k; \ell)\in\ncg 1; N-1\ncd^2$ with $k\neq\ell$, $\chi _{0;k}'\chi _{0;\ell}'=0$. 
\end{itemize}
Let $\chi _0'$ be the tensor product $\otimes_{k\in\ncg 1; N-1\ncd}\, \chi _{0;k}'$, that is the map 
\[(\R^3)^{N-1}\ni \ux'\ \donne\ \prod _{k\in\ncg 1; N-1\ncd}\, \chi _{0;k}'(x_{k}')\period\]
As above, we may assume that, for $k\in c(\cD )$, there is an open neighbourhood $\cV_{\neq k}'$ of $\hat{\ux}_k'$ such that $\cW_j\times\cV_{\neq k}'\subset\cV '$ and the support of $\otimes_{\ell\in\ncg 1; N-1\ncd\setminus\{k\}}\, \chi _{0;\ell}'$ is included in $\cV_{\neq k}'$. In particular, $\chi _0'\in\rC_c^\infty ((\R^3)^{N-1}; \R)$ and the support of $\chi _0'$ is included in $\cV'$.\\
The considered localised version of $\gamma _{N-1}$ is $\gamma _{N-1}(\chi _0\otimes\chi _0')$, that is the map 
\[((\R^3)^{N-1})^2\ni (\ux ; \ux')\ \donne \ \chi _0(\ux)\, \gamma _{N-1}(\ux ; \ux')\, \chi _0'(\ux ')\period\]
As explained above, it suffices to consider the localised version $\gamma _{N-1}^{(j)}(\chi _0\otimes\chi _0')$ of $\gamma _{N-1}^{(j)}$, that is the map 
\[\bigl((\R^3)^{N-1}\bigr)^2\ni (\ux ; \ux')\ \donne \ \chi _0(\ux)\, \gamma _{N-1}^{(j)}(\ux ; \ux')\, \chi _0'(\ux ')\comma\]
for $j\in\cD$. \\
Let us now fix $j\in\cD$. For the rest of Section~\ref{s:Fourier-matrice}, to simplify notation, we denote 
$\gamma _{N-1}^{(j)}$ by $\tilde\gamma$, the localised version of $\gamma _{N-1}^{(j)}$ by $\tilde\gamma _0$, i.e. $\tilde\gamma _0=\tilde\gamma (\chi _0\otimes\chi _0')$, $\tilde\chi _j$ by $\tilde\chi$, $c(j)$ by $k$, $\cV_{\neq j}$ by $\cV_{\neq}$, $\cV_{\neq k}'$ by $\cV_{\neq}'$, and $\cW_j$ by $\cW$. Therefore \eqref{eq:gamma-j} reads 
 \begin{equation}\label{eq:gamma}
 \tilde\gamma (\ux ;\, \ux ')\ =\ \int_{\R^3}\, |x_j\, -\, y|\; \tilde\varphi_j(\ux ; y)\; |x_k'\, -\, y|\; \overline{\tilde\varphi '}_k(\ux '; y)\; \tilde\chi (y)\, dy\period
 \end{equation}
According to \eqref{eq:fourier-g} and to the Fubini theorem, the Fourier transform $F$ of $\tilde\gamma _0$ is the map $F: (\R^3)^{2(N-1)}\ni (\uxi; \uxi ')\donne F(\uxi; \uxi ')$ where 
\begin{align}\label{eq:transf-fourier}
 &\ F(\uxi ; \uxi ')\\
 \ =&\ \int_{(\R^3)^{2N-1}}\, e^{-i(\uxi\cdot \ux\, +\, \uxi '\cdot \ux')}\, \chi _0(\ux)\; |x_j\, -\, y|\; \tilde\varphi_j(\ux ; y)\; \chi _0'(\ux')\; |x_k'\, -\, y|\; \overline{\tilde\varphi '}_k(\ux '; y)\; \tilde\chi (y)\; d\ux\, d\ux '\, dy\period\nonumber
\end{align}
It is convenient to write the variable $\ux\in (\R^3)^{N-1}$ as $(x_j; \ux_j)$ with $\ux_j:=(x_\ell)_{\ell\neq j}$. We also simplify the notation for the tensor product $\chi _0$ by setting 
\[\chi _j\bigl(\ux_j\bigr)\ :=\ \prod_{\ell\in\ncg 1; N-1\ncd\atop \ell\neq j}\, \chi _{0; \ell}(x_\ell)\]
if $N>2$, else $\chi _j=1$. Thus $\chi _0(\ux)=\chi _{0;j}(x_j)\, \chi_j(\ux_j)$. We perform the same simplifications for $\ux '$ and $\chi_0'$, replacing $j$ by $k$. \\
On $\cV\times\cW$, we have the special state decomposition $\psi (\ux; y)=\tilde\phi_j(\ux; y)+|x_j-y|\tilde\varphi_j(\ux; y)$ associated to the collision of the variables $x_j$ and $y$ at $\hat{x}_j$ (cf. Proposition~\ref{prop:décomp-anal}). As in Definition~\ref{def:valuation}, we write $(\ux; y)=(x_j; y; \ux_j)$ and $\tilde\varphi_j(\ux; y)=\varphi_j((x_j-y)/2; (x_j+y)/2; \ux_j)$. Recall that $n_j$ is the valuation of the map $\uz\donne\varphi_j(z_j; z_N; Z_{j;N})$ in the variable $z_j$ at $(0; \hat{x}_j; \hat{\ux}_j)$. Similarly, using the special state decomposition associated to the collision of the variables $x_k'$ and $y$ at $\hat{x}_k'=\hat{x}_j$, we can write $\tilde\varphi_k'(\ux '; y)=\varphi_k'((x_k'-y)/2; (x_k'+y)/2; \ux_k')$ and $n_k'$ is the valuation of the map $\uz\donne\varphi_k'(z_k; z_N; Z_{k;N})$ in the variable $z_k$ at $(0; \hat{x}_k'; \hat{\ux}_k')$.\\
To analyse the function $F$, we first rewrite it in the following form, using the functions $\varphi_j$ and $\varphi_k'$. 
\begin{lemma}\label{lm:change-var}
Let $\chi\in\rC_c^\infty(\R; \R)$ be such that, for all $x\in\R^3$ and $y\in S_{\tilde\chi}$, the support of $\tilde\chi$, $\chi (|x|)=1$ if $\chi _{0;j}(x+y)\neq 0$ and also $\chi (|x|)=1$ if $\chi _{0;k}'(x+y)\neq 0$. Then $\chi (|\cdot|)=1$ near $0$ in $\R^3$. For all $(\uxi ; \uxi ')\in (\R^3)^{2(N-1)}$, 
\begin{align}
 F(\uxi ; \uxi ')\ =&\ \int_{(\R^3)^{2N-1}}\; e^{-i\,(\xi_j\cdot x_j\, +\, \xi_k'\cdot x_k')}\; e^{-i(\xi _j+\xi'_k)\cdot y}\; e^{-i\,(\uxi _j\cdot\ux_j\, +\, \uxi_k'\cdot\ux_k')}\label{eq:fourier-tilde-varphi}\\
 &\ \hspace{.8cm} |x_j|\; \varphi _j\bigl(x_j/2;\, x_j/2+y\, ;\, \ux _j\bigr)\; \chi \bigl(|x_j|\bigr)\ |x_k'|\; \overline{\varphi '}_k\bigl(x_k'/2;\, x_k'/2+y\, ;\, \ux _k'\bigr)\; \chi \bigl(|x_k'|\bigr)\nonumber\\
 &\ \hspace{.8cm} \chi _{0; j}(x_j+y)\; \chi _{0; k}'(x_k'+y)\; \chi _j\bigl(\ux_j\bigr)\; \chi _k'\bigl(\ux_k'\bigr)\; \tilde\chi (y)\;  dx_j\, dx_k'\, d\ux _j\, d\ux _k'\, dy\comma\nonumber
\end{align}
where $\uxi :=(\xi_j; \uxi_j)$ and $\uxi ':=(\xi_j'; \uxi_j')$ with $\uxi_j:=(\xi_\ell)_{\ell\neq j}$ and $\uxi_k':=(\xi_\ell ')_{\ell\neq k}$.
\end{lemma}
\Pf As above, we write $(\ux; y)=(x_j; y; \ux_j)$ and $(\ux '; y)=(x_k'; y; \ux_k')$ in \eqref{eq:transf-fourier}. For $y\in S_{\tilde\chi}$, $\chi _{0;j}(x+y)\neq 0$ if $x$ is close to zero. Thus $\chi (|x|)=1$ for such $x$. Similarly, for $y\in S_{\tilde\chi}$, $\chi _{0;k}'(x+y)\neq 0$ if $x$ is close to zero. Thus $\chi (|x|)=1$ for such $x$, as well.\\
For fixed $y$, we make the change of variables $\tilde{x}=x_j-y$ and $\tilde{x}'=x_k'-y$ in the $(x_j; x_k')$-integral in \eqref{eq:transf-fourier} and rename the new variables $(\tilde{x}; \tilde{x}')$ as $(x_j; x_k')$, to arrive at 
\begin{align*}
 F(\uxi ; \uxi ')\ =&\ \int_{(\R^3)^{2N-1}}\; e^{-i\,(\xi_j\cdot x_j\, +\, \xi_k'\cdot x_k')}\; e^{-i(\xi _j+\xi'_k)\cdot y}\; e^{-i\,(\uxi _j\cdot\ux_j\, +\, \uxi_k'\cdot\ux_k')}\\
 &\ \hspace{.8cm} |x_j|\; \tilde\varphi _j\bigl(x_j+y; y; \ux _j\bigr)\; |x_k'|\; \overline{\tilde\varphi '}_k\bigl(x_k'+y; y; \ux_k'\bigr)\; \\
 &\ \hspace{.8cm} \chi _{0; j}(x_j+y)\; \chi _{0; k}'(x_k'+y)\; \chi _j\bigl(\ux_j\bigr)\; \chi _k'\bigl(\ux_k'\bigr)\; \tilde\chi (y)\;  dx_j\, dx_k'\, d\ux _j\, d\ux _k'\, dy\period
\end{align*}
We can write $\tilde\varphi _j(x_j+y; y; \ux _j)=\varphi _j(x_j/2; x_j/2+y; \ux _j)$, for $y\in S_{\tilde\chi}$ and $(x_j+y)\in S_{\chi _{0;j}}$, and $\tilde\varphi _k'(x_k'+y; y; \ux _k')=\varphi _k'(x_k'/2; x_k'/2+y; \ux _k')$, 
for $y\in S_{\tilde\chi}$ and $(x_k'+y)\in S_{\chi _{0;k}'}$. By the properties of the cut-off function $\chi$, we may insert $\chi (|x_j|)$ and $\chi (|x_k'|)$ into the integral without changing it, yielding \eqref{eq:fourier-tilde-varphi}. $\cqfd$

Recall that $\hat{x}_j=\hat{x}_k'$. Let $\cR_0$ be the range of the map $\cW^2\ni (x; y)\donne ((x-y)/2; (x+y)/2)$. It is an 
open neighbourhood of $(0; \hat{x}_j)=(0; \hat{x}_k')$. Let 
\[\cS\ :=\ \bigl\{(s; t)\in (\R^3)^2\, ;\ (s/2; s/2+t)\in\cR_0\bigr\}\period\]
On the support of the integrand in \eqref{eq:fourier-tilde-varphi}, we have $(x_j/2; x_j/2+y)\in\cR_0$, that is $(x_j; y)\in\cS$, 
and $(x_k'/2; x_k'/2+y)\in\cR_0$, that is $(x_k'; y)\in\cS$. It was shown in \cite{jn} (cf. the proof of the claim (24) in the appendix) that, for almost all $y\in S_{\tilde\chi}$,  $n_j$ is the valuation of the map 
\[\cS\times\cV_{\neq}\ni (x_j; y; \ux_j)\ \donne\ \varphi _j\bigl(x_j/2;\, x_j/2+y\, ;\, \ux _j\bigr)\]
at $0$ in the variable $x_j$ and that $n_k'$ is the valuation of the map 
\[\cS\times\cV_{\neq}'\ni (x_k'; y; \ux_k')\ \donne\ \varphi _k'\bigl(x_k'/2;\,  x_k'/2+y\, ;\, \ux _k'\bigr)\]
at $0$ in the variable $x_k'$. \\
Let $m$ be an integer larger than $\max (n_j; n_k')$. Then, for $(x_j/2; x_j/2+y; \ux _j)\in\cR_0\times\cV_{\neq}$, 
\begin{equation}\label{eq:dév-varphi-j}
\varphi _j\bigl(x_j/2;\, x_j/2+y\, ;\, \ux _j\bigr)\ =\ \sum _{\alpha\in\N^3\atop n_j\leq |\alpha |<m}\, \varphi_\alpha (y; \ux _j)\; x_j^\alpha\, +\, \sum _{\alpha\in\N^3\atop |\alpha |=m}\, \hat{\varphi}_{\alpha}\bigl(x_j;\, y\, ;\, \ux _j\bigr)\; x_j^\alpha\comma 
\end{equation}
where the functions $\hat{\varphi}_{\alpha}$ are the sum of a power series on $\cS\times\cV_{\neq}$ and the functions $\varphi_\alpha$ are the sum of a power series on $\cW\times\cV_{\neq}$. By definition of $n_j$, the functions $\varphi_\alpha$ for $|\alpha |=n_j$, are not all zero. \\
Similarly, we can write, for $(x_k'; x_k'+2y; \ux _k')\in\cR_0\times\cV_{\neq}'$, 
\begin{equation}\label{eq:dév-varphi'-k}
\varphi _k'\bigl(x_k'/2;\,  x_k'/2+y\, ;\, \ux _k'\bigr)\ =\ \sum _{\alpha\in\N^3\atop n_k'\leq |\alpha |<m}\, \varphi_\alpha '(y; \ux _k')\; (x_k')^\alpha\, +\, \sum _{\alpha\in\N^3\atop |\alpha |=m}\, \hat{\varphi}_{\alpha}'\bigl(x_k';\, y\, ;\, \ux _k'\bigr)\; (x_k')^\alpha\comma 
\end{equation}
where the functions $\hat{\varphi}_{\alpha}'$ are the sum of a power series near on $\cS\times\cV_{\neq}'$ and the functions $\varphi_\alpha '$ are the sum of a power series on $\cW\times\cV_{\neq}'$.
By definition of $n_k'$, the functions $\varphi_\alpha '$ for $|\alpha |=n_k'$, are not all zero. \\
Now we want to insert the formulae \eqref{eq:dév-varphi-j} and \eqref{eq:dév-varphi'-k} into \eqref{eq:fourier-tilde-varphi}. To control the remainders, we shall use Lemma~\ref{lm:inté-parties} and Lemma~\ref{lm:fourier-norme-localisée}. \\
For all $\alpha\in\N^3$, we denote by $F_\alpha$ the Fourier transform of the continuous, compactly supported function $\R^3\ni x\donne |x|\, x^\alpha\, \chi(|x|)$. We observe that $F_\alpha=(-i\partial)^\alpha F_0$, where $F_0$ is the Fourier transform of the continuous, compactly supported function $\R^3\ni x\donne |x|\, \chi(|x|)$ and to which one can apply Lemma~\ref{lm:fourier-norme-localisée}. This yields 
\begin{lemma}\label{lm:comportement-F-alpha}
Let $\alpha\in\N^3$. There exists a smooth function $G_\alpha : \R^3\setminus\{0\}\dans\C$ satisfying the following properties. 
\begin{equation}\label{eq:decomp-F-alpha}
 \forall\, \xi\in \R^3\setminus\{0\}\comma\hspace{.4cm}F_\alpha (\xi)\ =\ -8\pi\, (-i)^{|\alpha|}\, \left(\partial^\alpha\, |\cdot|^{-4}\right)(\xi)\, +\, G_\alpha (\xi)\hspace{.4cm}\mbox{and}
\end{equation}
\begin{equation}\label{eq:comportement-G-alpha}
\forall\, \gamma\in\N^3\comma\ \forall\, k\in\N\comma\ \exists\, C_{\gamma ; k}^{(\alpha)}>0\, ;\ \forall\, \xi\in \R^3\setminus\{0\}\comma\ \bigl|\partial^\gamma G_\alpha(\xi)\bigr|\ \leq\ C_{\gamma ; k}^{(\alpha)}\;  |\xi |^{-k-|\gamma|}\period
\end{equation}
In particular, we have 
\begin{equation}\label{eq:comportement-F-alpha}
\forall\, \gamma\in\N^3\comma\ \exists\, C_\gamma^{(\alpha)}>0\, ;\ \forall\, \xi\in \R^3\setminus\{0\}\comma\ \bigl|\partial^\gamma F_\alpha(\xi)\bigr|\ \leq\ C_\gamma^{(\alpha)}\;  |\xi |^{-4-|\alpha|-|\gamma|}\period
\end{equation}
\end{lemma}
We can check by induction that, for all $\alpha\in\N^3$, there exists a polynomial $P_\alpha$ on $\R^3$, that has real coefficients and is homogeneous of degree $|\alpha|$, such that 
\begin{equation}\label{eq:dérivée-alpha-norme-4}
\forall\, \eta\in\R^3\setminus\{0\}\comma\ \Bigl(\partial^\alpha |\cdot|^{-4}\Bigr)(\eta)\ =\ \frac{P_\alpha (\eta)}{|\eta|^{4+2|\alpha|}}\period
\end{equation}
For further purpose, we state the following 
\begin{lemma}\label{lm:famille-libre}
Let $n\in\N$. The functions $F_\alpha$ for $|\alpha|=n$ are linearly independent. So are the functions $P_\alpha$ for $|\alpha|=n$.
\end{lemma}
\Pf See the Appendix. $\cqfd$

Coming back to \eqref{eq:fourier-tilde-varphi}, this leads to the 
\begin{proposition}\label{prop:dév-fourier-gamma}
Let $m$ be any integer larger than $\max (n_j; n_k'; 3)$. Then there exist a family $(a_\alpha)_{n_j\leq |\alpha |<m}$ of smooth functions, defined near $\hat{x}$ in $\R^{3(N-1)}$, a family $(a_\alpha ')_{n_k'\leq |\alpha |<m}$ of smooth functions, defined near $\hat{x}'$ in $\R^{3(N-1)}$, and a smooth function $R_m$ on $(\R^3)^{2(N-1)}$, such that, for all $(\uxi ; \uxi')\in (\R^3)^{2(N-1)}$, 
\begin{equation}\label{eq:décomp-fourier-gamma-m}
 F(\uxi ; \uxi')\ =\ \sum _{n_j\leq |\alpha|<m\atop n_k'\leq |\alpha '|<m}\, F_\alpha (\xi _j)\, F_{\alpha '}(\xi _k')\, A_{\alpha ; \alpha '}\bigl(\uxi ; \uxi '\bigr)\ +\ R_m(\uxi ; \uxi')\comma
\end{equation}
where 
\begin{equation}\label{eq:fourier-jky}
 A_{\alpha ; \alpha '}\bigl(\uxi ; \uxi '\bigr)\ =\ \int_{\R^3}\, e^{-i(\xi _j+\xi _k')\cdot y}\, B_{\alpha; \alpha '}\bigl(\uxi _j; \uxi _k'; y\bigr)\, \tilde\chi (y)\, dy
\end{equation}
and $B_{\alpha; \alpha '}\bigl(\uxi _j; \uxi _k'; y\bigr)$ is the integral
\begin{equation}\label{eq:fourier-jk}
 \int_{\R^{3(2N-4)}}\, e^{-i\,(\uxi _j\cdot\ux_j\, +\, \uxi_k'\cdot\ux_k')}\; a_\alpha (y; \ux_j)\; \overline{a'}_{\alpha '}(y; \ux_k')\; \chi _j\bigl(\ux_j\bigr)\; \chi _k'\bigl(\ux_k'\bigr)\; d\ux _j\, d\ux _k'\period
\end{equation}
The functions $A_{\alpha ; \alpha '}$ and $B_{\alpha; \alpha '}$ are smooth. \\
For $\alpha\in\N^3$ with $|\alpha|=n_j$, we have $a_\alpha (y; \ux_j)=\varphi _\alpha (y; \ux _j)\chi _{0; j}(y)$, where the function $\varphi _\alpha$ appears in the formula \eqref{eq:dév-varphi-j}, and, for $\alpha '\in\N^3$ with $|\alpha '|=n_k'$, we have $a_{\alpha '}'(y; \ux_k')=\varphi _{\alpha '}'(y; \ux _k')\chi _{0; k}'(y)$, where the function $\varphi _{\alpha '}'$ appears in the formula \eqref{eq:dév-varphi'-k}.\\
Moreover, we have the following estimates. 
\begin{equation}\label{eq:esti-reste}
 \forall\, q\in\N\comma\hspace{.4cm}\sup _{(\uxi ;\, \uxi')\in (\R^3)^{2(N-1)}}\; \langle \xi_j\rangle^m\, \langle \xi_k'\rangle^m\, \bigl\langle \uxi_j\bigr\rangle^q\, \bigl\langle \uxi_k'\bigr\rangle^q\, \bigl|R_m(\uxi ; \uxi')\bigr|\ <\ +\infty\period
\end{equation}
For $\alpha\in\N^3$ with $m>|\alpha|\geq n_j$ and $\alpha '\in\N^3$ with $m>|\alpha '|\geq n_k'$, for all $q\in\N$, 
\begin{equation}\label{eq:esti-termes}
 \sup _{(\uxi ;\, \uxi')\in (\R^3)^{2(N-1)}}\; \langle \xi_j\rangle^{4+|\alpha|}\, \langle \xi_k'\rangle^{4+|\alpha '|}\, \bigl\langle \uxi_j\bigr\rangle^q\, \bigl\langle \uxi_k'\bigr\rangle^q\, \bigl|F_\alpha (\xi _j)\, F_{\alpha '}(\xi _k')\, A_{\alpha ; \alpha '}\bigl(\uxi ; \uxi '\bigr)\bigr|\ <\ +\infty\period
\end{equation}
\end{proposition}
\Pf Let $y\in S_{\tilde\chi}$ be fixed. We write a Taylor formula for $\chi _{0; j}$ at fixed $y$ with exact remainder as an integral: 
\[\chi _{0; j}(x_j+y)\ =\ \sum _{|\delta |<m-n_j}\, \frac{\partial^{\delta }\chi_{0; j}(y)}{\delta !}\; x_j^\delta \, +\, \sum _{|\delta |=m-n_j}\, \hat{\chi}_{\delta}(x_j;\, y)\; x_j^\delta\comma\]
where the functions $\hat{\chi}_{\delta}$ are smooth near $(0; \hat{x}_j)$. Using this formula together with \eqref{eq:dév-varphi-j}, we get the following expansion on $\cS\times\cV_{\neq}$, 
\[\varphi _j\bigl(x_j/2;\, x_j/2+y\, ;\, \ux _j\bigr)\, \chi _{0; j}(x_j+y)\ =\ \sum _{n_j\leq |\alpha |<m}\, a_\alpha (y; \ux_j)\; x_j^\alpha\, +\, \sum _{m\leq |\delta |\leq 2m-n_j}\, r_\delta(x_j; y; \ux_j)\; x_j^\delta\comma\]
for some smooth functions $a_\alpha$ and $r_\delta$. For $|\alpha |=n_j$, $a_\alpha (y; \ux_j)=\varphi _\alpha (y; \ux _j)\chi _{0; j}(y)$. 
In the last formula, denote by $p(x_j; y; \ux_j)$ the first sum and by $r(x_j; y; \ux_j)$ the second one.\\
Similarly, using \eqref{eq:dév-varphi'-k}, we can write on $\cS\times\cV_{\neq}'$
\[\varphi _k'\bigl(x_k'/2;\, x_k'/2+y\, ;\, \ux _k'\bigr)\chi _{0; k}'(x_k'+y)\ =\ p'\bigl(x_k'; y; \ux_k'\bigr)\, +\, r'\bigl(x_k'; y; \ux_k'\bigr)\period\]
Inserting these expansions into \eqref{eq:fourier-tilde-varphi}, we obtain 
\begin{align}
 F(\uxi ; \uxi ')\ =&\ \int_{(\R^3)^{2N-1}}\; e^{-i\,(\xi_j\cdot x_j\, +\, \xi_k'\cdot x_k')}\; e^{-i(\xi _j+\xi'_k)\cdot y}\; e^{-i\,(\uxi _j\cdot\ux_j\, +\, \uxi_k'\cdot\ux_k')}\label{eq:expression-F}\\
 &\ \hspace{.8cm} |x_j|\; \chi \bigl(|x_j|\bigr)\ |x_k'|\; \chi \bigl(|x_k'|\bigr)\; \bigl(p(x_j; y; \ux_j)\, +\, r(x_j; y; \ux_j)\bigr)\; \tilde\chi (y)\nonumber\\
 &\ \hspace{.8cm} \bigl(\overline{p'}(x_k'; y; \ux_k')\, +\, \overline{r'}(x_k'; y; \ux_k')\bigr)\; \chi _j\bigl(\ux_j\bigr)\; \chi _k'\bigl(\ux_k'\bigr)\; dx_j\, dx_k'\, d\ux _j\, d\ux _k'\, dy\nonumber
\end{align}
and set $R_m(\uxi ; \uxi ')$ as 
\begin{align*}
 F(\uxi ; \uxi ')&\ -\, \int_{(\R^3)^{2N-1}}\; e^{-i\,(\xi_j\cdot x_j\, +\, \xi_k'\cdot x_k')}\; e^{-i(\xi _j+\xi'_k)\cdot y}\; e^{-i\,(\uxi _j\cdot\ux_j\, +\, \uxi_k'\cdot\ux_k')}\\
 &\ \hspace{1.2cm} |x_j|\; \chi \bigl(|x_j|\bigr)\ |x_k'|\; \chi \bigl(|x_k'|\bigr)\; p(x_j; y; \ux_j)\; \tilde\chi (y) \\
 &\ \hspace{1.2cm} \overline{p'}(x_k'; y; \ux_k')\; \chi _j\bigl(\ux_j\bigr)\; \chi _k'\bigl(\ux_k'\bigr)\; dx_j\, dx_k'\, d\ux _j\, d\ux _k'\, dy\period
\end{align*}
By Fubini theorem, we have 
\begin{align*}
 &\ F(\uxi ; \uxi ')\, -\ R_m(\uxi ; \uxi ')\\
 \ =&\ \sum _{n_j\leq |\alpha |<m\atop n_k'\leq |\alpha '|<m}\, \int_{\R^6}\; e^{-i\,(\xi_j\cdot x_j\, +\, \xi_k'\cdot x_k')}\; x_j^\alpha\; |x_j|\; \chi \bigl(|x_j|\bigr)\ (x_k')^{\alpha '}\; |x_k'|\; \chi \bigl(|x_k'|\bigr)\; dx_j\, dx_k'\\
 &\ \hspace{2.1cm} \times\ \int_{\R^3}\, e^{-i(\xi _j+\xi'_k)\cdot y}\; B_{\alpha ; \alpha '}(\uxi _j; \uxi_k'; y)\; \tilde\chi (y)\, dy\comma
\end{align*}
where $B_{\alpha ; \alpha '}(\uxi _j; \uxi_k'; y)$ is given by \eqref{eq:fourier-jk}, 
yielding \eqref{eq:décomp-fourier-gamma-m}. \\
For such $(\alpha ; \alpha ')$, we can use \eqref{eq:dérivée-exp-crochet} in \eqref{eq:fourier-jk} to get, for all $q\in\N$, 
\[\forall\, (\uxi _j; \uxi_k'; y)\in \bigl(\R^{3(N-2)}\bigr)^2\times S_{\tilde\chi}\comma\hspace{.4cm}\bigl|B_{\alpha ; \alpha '}(\uxi _j; \uxi_k'; y)\bigr|\ \leq\ C_q\, \bigl\langle \uxi_j\bigr\rangle^{-q}\, \bigl\langle \uxi_k'\bigr\rangle^{-q}\, \]
for some $(\uxi _j; \uxi_k'; y)$-independent constant $C_q$. Combining this estimate with \eqref{eq:fourier-jky} and \eqref{eq:comportement-F-alpha}, we derive \eqref{eq:esti-termes}. \\
The rest $R_m(\uxi ; \uxi ')$ is given by 
\begin{align*}
 &\sum_{s\in\{p; r\}\, ;\, s'\in\{p'; r'\}\, ;\atop (s; s')\neq (p; p')}\int_{(\R^3)^{2N-1}}\; e^{-i\,(\xi_j\cdot x_j\, +\, \xi_k'\cdot x_k')}\; e^{-i(\xi _j+\xi'_k)\cdot y}\; e^{-i\,(\uxi _j\cdot\ux_j\, +\, \uxi_k'\cdot\ux_k')}\\
 &\ \hspace{3.2cm} |x_j|\; \chi \bigl(|x_j|\bigr)\ |x_k'|\; \chi \bigl(|x_k'|\bigr)\; s(x_j; y; \ux_j)\; \tilde\chi (y) \\
 &\ \hspace{3.2cm} \overline{s'}(x_k'; y; \ux_k')\; \chi _j\bigl(\ux_j\bigr)\; \chi _k'\bigl(\ux_k'\bigr)\; dx_j\, dx_k'\, d\ux _j\, d\ux _k'\, dy\period
\end{align*}
Let $q\in\N$. Take a term in the above sum that contains the function $r$. By Lemma~\ref{lm:régu-valuation}, we know that the map $x_j\donne |x_j|\chi (|x_j|)r(x_j; y; \ux_j)$ belongs to the class $\rC^m$. Thus we can apply precisely $m$ times the identity \eqref{eq:dérivée-exp-crochet} with $(x; \xi)$ replaced by $(x_j; \xi _j)$ and integrations by parts in the $x_j$-integral, we use $q$ times \eqref{eq:dérivée-exp-crochet} with $(x; \xi)$ replaced by $(\ux _j; \uxi _j)$ and integrations by parts in the $\ux_j$-integral, and we use $q$ times \eqref{eq:dérivée-exp-crochet} with $(x; \xi)$ replaced by $(\ux _k'; \uxi _k')$ and integrations by parts in the $\ux_k'$-integral, to get 
\begin{equation}\label{eq:borne-reste}
 \langle \xi_j\rangle^m\, \langle \xi_k'\rangle^m\, \langle \uxi_j\rangle^q\, \langle \uxi_k'\rangle^q\, \bigl|R_m(\uxi ; \uxi ')\bigr|\ \leq\ C\comma
\end{equation}
for some $(\uxi ; \uxi ')$-independent constant $C$. The remaining term contains the function $r'$. By Lemma~\ref{lm:régu-valuation}, the map $x_k'\donne |x_k'|\chi (|x_k'|)\overline{r}'(x_j; y; \ux_j)$ belongs to the class $\rC^m$. We use $m$ times the identity \eqref{eq:dérivée-exp-crochet} with $(x; \xi)$ replaced by $(x_k'; \xi _k')$ and integrations by parts in the $x_k'$-integral, we use $q$ times \eqref{eq:dérivée-exp-crochet} with $(x; \xi)$ replaced by $(\ux _j; \uxi _j)$ and integrations by parts in the $\ux_j$-integral, and we use $q$ times \eqref{eq:dérivée-exp-crochet} with $(x; \xi)$ replaced by $(\ux _k'; \uxi _k')$ and integrations by parts in the $\ux_k'$-integral, to get the estimate \eqref{eq:borne-reste}, for a possibly different, $(\uxi ; \uxi ')$-independent constant $C$. This yields \eqref{eq:esti-reste}. $\cqfd$

The estimates \eqref{eq:esti-reste} and \eqref{eq:esti-termes} ensure, by the Fubini theorem, that each term in \eqref{eq:décomp-fourier-gamma-m} is integrable over $(\R^3)^{2(N-1)}$. This allows us to apply \eqref{eq:fourier-inv-g-bis}, with $F_g$ replaced by these terms. In particular, $\tilde\gamma _0$ is the inverse Fourier transform of $F$. It is convenient to introduce the map $\gamma _0 : \R^{3(N-1)}\dans\C$ that is defined by 
\begin{equation}\label{eq:def-gamma-0}
 \gamma _0 \bigl(X; X';\, \ux_j;\, \ux _k'\bigr)\ =\ \tilde\gamma _0\bigl(X/2+X';\, \ux_j;\,  X'-X/2;\, \ux _k'\bigr)\period
\end{equation}
Of course, we can recover $\tilde\gamma_0$ from $\gamma _0$ by 
\begin{equation}\label{eq:tilde-gamma-0-gamma-0}
\tilde\gamma _0\bigl(x_j;\, \ux_j;\,  x_k';\, \ux _k'\bigr)\ =\ \gamma _0 \bigl(x_j-x_k'; (x_j+x_k')/2;\, \ux_j;\, \ux _k'\bigr)\period
\end{equation}
In particular, $\tilde\gamma_0$ and $\gamma _0$ have the same regularity. Let us denote by $\cR$ the range of the map $\cW^2\ni(x; y)\donne (x-y; (x+y)/2)$. It is a neighbourhood of $(0; \hat{x}_j)=(0; \hat{x}_k')$. 
\begin{proposition}\label{prop:dév-gamma}
Let $m$ be any integer larger than $4+n_j+n_k'$. Then there exists a function $\rR_m : \cR\times\cV_{\neq}\times\cV_{\neq}'\dans\C$, that belongs to the class $\rC^m$, such that, for all $(X; X';\, \ux_j;\, \ux _k')\in\cR\times\cV_{\neq}\times\cV_{\neq}'$, 
\begin{align}
&\ \gamma _0 \bigl(X; X';\, \ux_j;\, \ux _k'\bigr)\ -\ \rR_m\bigl(X; X';\, \ux_j;\, \ux _k'\bigr)\label{eq:dev-gamma-1}\\
\ =&\ \chi _j\bigl(\ux_j\bigr)\; \chi _k'\bigl(\ux_k'\bigr)\; \sum _{n_j\leq |\alpha |\, ,\  n_k'\leq |\alpha '|\atop 4+|\alpha|+|\alpha '|<m}\ \sum _{4+|\alpha|+|\beta|+|\alpha '|+|\beta '|<m}\, \frac{1}{\beta !\, \beta '!}\nonumber\\
&\hspace{4.5cm}\times
\Bigl(\frac{\partial _y}{2}\Bigr)^{\beta +\beta '}\Bigl(\tilde\chi (y)\, a_\alpha (y; \ux_j)\, \overline{a'}_{\alpha '}(y; \ux_k')\Bigr)_{|y=X'}\nonumber\\
&\hspace{4.5cm}\times\, (2\pi)^{-3}\int_{\R^3}\, e^{iX\cdot \eta}\, F_{\alpha +\beta}(\eta)\, F_{\alpha '+\beta '}(-\eta)\, d\eta\period\nonumber
\end{align}
Moreover, for fixed $(\alpha ; \alpha '; \beta ; \beta ')$, the term 
\begin{align}
 &\frac{1}{\beta !\, \beta '!}\ \Bigl(\frac{\partial _y}{2}\Bigr)^{\beta +\beta '}\Bigl(\tilde\chi (y)\, a_\alpha (y; \ux_j)\, \overline{a'}_{\alpha '}(y; \ux_k')\Bigr)_{|y=X'}\label{eq:term-expansion}\\
 &\hspace{4.5cm}\times\, (2\pi)^{-3}\int_{\R^3}\, e^{iX\cdot \eta}\, F_{\alpha +\beta}(\eta)\, F_{\alpha '+\beta '}(-\eta)\, d\eta\nonumber
\end{align}
belongs to the class $\rC^{4+|\alpha |+|\beta|+|\alpha '|+|\beta '|}$.
\end{proposition}
\Pf We start with formula \eqref{eq:décomp-fourier-gamma-m} with $m$ replaced by $m+4$. Denote by $F^A(\uxi; \uxi ')$ the sum on the r.h.s. of this formula. Thanks to the estimate \eqref{eq:esti-reste} and Lemma~\ref{lm:inté-parties}, the inverse Fourier transform of $R_{m+4}$ belongs to the class $\rC^m$. For two functions $f$ and $g$, we write $f\sim g$ if $f-g$ belongs to the class $\rC^m$. \\
Now, we apply the inverse Fourier transform \eqref{eq:fourier-inv-g} to $F^A$. For $(\ux; \ux')\in (\R^3)^{2(N-1)}$, 
\begin{align}
&\ (2\pi)^{-6(N-1)}\, \int_{(\R^3)^{2(N-1)}}\, e^{i(\ux\cdot\uxi +\ux'\cdot\uxi ')}\, F^A(\uxi ; \uxi ')\, d\uxi\, d\uxi '\label{eq:fourier-inv-F-A}\\
\ =&\ (2\pi)^{-6}\, \chi _j\bigl(\ux_j\bigr)\; \chi _k'\bigl(\ux_k'\bigr)\; \sum _{n_j\leq |\alpha|<m+4\atop n_k'\leq |\alpha '|<m+4}\, \int_{\R^6}\, e^{i(x_j\cdot\xi _j +x'_k\cdot\xi _k')}\, F_\alpha (\xi _j)\, F_{\alpha '}(\xi _k')\, \int_{\R^3}\, e^{-i(\xi _j+\xi _k')\cdot y}\, \tilde\chi (y)\nonumber\\ 
&\hspace{7.5cm}\times\, a_\alpha (y; \ux_j)\, \overline{a'}_{\alpha '}(y; \ux_k')\; dy\, d\xi _j\, d\xi _k'\period\nonumber
\end{align}
Actually, this is zero unless $(x_j; x_k'; \ux_j; \ux_k')\in\cW^2\times\cV_{\neq}\times\cV_{\neq}'$.
In the latter case, let us denote by $\tilde{C}_{\alpha ; \alpha '}(x_j; x_k'; \ux _j; \ux _k')$ the terms appearing in the sum in \eqref{eq:fourier-inv-F-A} and set 
\[C_{\alpha ; \alpha '}\bigl(X; X'; \ux _j; \ux _k'\bigr)\ :=\ \tilde{C}_{\alpha ; \alpha '}\bigl(X'+X/2; X'-X/2; \ux _j; \ux _k'\bigr)\period\]
where $(X; X'; \ux _j; \ux _k')\in\cR\times\cV_{\neq}\times\cV_{\neq}'$.
Making the change of variables $\eta :=(\xi _j-\xi _k')/2$ and $\eta ':=(\xi _j+\xi _k')/2$, we obtain 
\begin{align*}
 \tilde{C}_{\alpha ; \alpha '}(x_j; x_k'; \ux _j; \ux _k)\ =&\ \int_{\R^6}\, e^{i(x_j\cdot (\eta +\eta ')+x'_k\cdot (\eta '-\eta))}\, F_\alpha \bigl(\eta +\eta '\bigr)\, F_{\alpha '}\bigl(\eta '-\eta\bigr)\\
 &\hspace{3.5cm}\times\, \int_{\R^3}\, e^{-2i\eta '\cdot y}\, \tilde\chi (y)\, a_\alpha (y; \ux_j)\, \overline{a'}_{\alpha '}(y; \ux_k')\; dy\, d\eta\, d\eta '
\end{align*}
and, setting $X:=x_j-x_k'$ and $X':=(x_j+x_k')/2$, we get 
\begin{align*}
 C_{\alpha ; \alpha '}(X; X'; \ux _j; \ux _k)\ =&\ \int_{\R^6}\, e^{i(X\cdot\eta + 2X'\cdot\eta ')}\, F_\alpha \bigl(\eta +\eta '\bigr)\, F_{\alpha '}\bigl(\eta '-\eta\bigr)\\
 &\hspace{3.5cm}\times\, \int_{\R^3}\, e^{-2i\eta '\cdot y}\, \tilde\chi (y)\, a_\alpha (y; \ux_j)\, \overline{a'}_{\alpha '}(y; \ux_k')\; dy\, d\eta\, d\eta '\period
\end{align*}
Using the identity \eqref{eq:dérivée-exp-crochet} with $(x; \xi)$ replaced by $(2y; \eta ')$, 
we see by integration by parts that, for all $q\in\N$, there exists $C_q>0$ such that, for $\ux_j$ in the support of $\chi _j$ and $\ux_k'$ in the support of $\chi _k'$, 
\begin{equation}\label{eq:int-parties-y}
\left|\langle \eta '\rangle^q\, \int_{\R^3}\, e^{-2i\eta '\cdot y}\, \tilde\chi (y)\, a_\alpha (y; \ux_j)\, \overline{a'}_{\alpha '}(y; \ux_k')\; dy\right|\ \leq \ C_q\period
\end{equation}
Using the boundedness of the functions $F_\alpha$ (cf. \eqref{eq:comportement-F-alpha}) and the estimate  \eqref{eq:int-parties-y}, we see that contribution to the integral of the region $\{\eta ;\, |\eta|<1\}$ is a smooth function (by standard derivation under the integral sign). This holds also true for the contribution to the integral of the region $\{(\eta ; \eta ');\, |\eta|\geq 1\comma\, |\eta|<2|\eta '|\}$. Thus $C_{\alpha ; \alpha '}\sim C_{\alpha ; \alpha '}^A$ where 
\begin{align*}
 C_{\alpha ; \alpha '}^A(X; X'; \ux _j; \ux _k)\ =&\ \int_{|\eta|\geq 1\atop 2|\eta '|\leq |\eta |}\, e^{i(X\cdot\eta + 2X'\cdot\eta ')}\, F_\alpha (\eta +\eta ')\, F_{\alpha '}(\eta '-\eta)\, \int_{\R^3}\, e^{-2i\eta '\cdot y}\, \tilde\chi (y)\\
 &\hspace{5.5cm}\times\, a_\alpha (y; \ux_j)\, \overline{a'}_{\alpha '}(y; \ux_k')\; dy\, d\eta\, d\eta '\period
\end{align*}
Now we use the following Taylor formula with exact remainder as an integral. For $\epsilon\in\{-1; 1\}$, $Q\in\N^\ast$, and $\alpha\in\N^3$, $F_\alpha (\eta '+\epsilon\eta)=F_\alpha ^A(\epsilon\eta ; \eta ')+F_\alpha ^R(\epsilon\eta ; \eta ')$ with 
\[F_\alpha ^A(\epsilon\eta ; \eta ')\ =\ \sum _{q=0}^{Q-1}\, \frac{1}{q!}\, \Bigl(\bigl(\eta '\cdot\nabla)^qF_\alpha\Bigr)(\epsilon\eta)\]
and 
\[F_\alpha ^R(\epsilon\eta ; \eta ')\ =\ \int_0^1\, \Bigl(\bigl(\eta '\cdot\nabla)^QF_\alpha\Bigr)(t\eta '+\epsilon\eta)\, \frac{(1-t)^{Q-1}}{(Q-1)!}\, dt\period\]
For $\epsilon\in\{-1; 1\}$, $|\eta|\geq 1$, $|\eta|\geq 2|\eta '|$, and $t\in [0; 1]$, $|t\eta '+\epsilon\eta|\geq |\eta|/2$ and, thanks to \eqref{eq:comportement-F-alpha}, there exist $C, C_Q, C_Q'>0$ such that 
\[\bigl|F_\alpha ^A(\epsilon\eta ; \eta ')\bigr|\ \leq\ C\, |\eta|^{-4-|\alpha|}\, |\eta '|^{Q}\comma\hspace{.4cm}\bigl|F_\alpha ^R(\epsilon\eta ; \eta ')\bigr|\ \leq\ C_Q\, |\eta|^{-4-|\alpha|-Q}\, |\eta '|^{Q}\comma\]
and 
\[\bigl|F_\alpha (\eta '+\eta)F_{\alpha '} ^R(-\eta ; \eta ')\bigr|\ \leq\ C_Q'\, |\eta|^{-8-|\alpha|-|\alpha '|-Q}\, |\eta '|^{2Q}\period\]
Set $Q=m-|\alpha|-|\alpha '|-4$ if $|\alpha|+|\alpha '|+4<m$, else $Q=1$. 
Using \eqref{eq:int-parties-y} and Lemma~\ref{lm:inté-parties}, we have $C_{\alpha ; \alpha '}^A\sim C_{\alpha ; \alpha '}^{A_1}$ where 
\begin{align*}
 C_{\alpha ; \alpha '}^{A_1}(X; X'; \ux _j; \ux _k)\ =&\ \int_{|\eta|\geq 1\atop 2|\eta '|\leq |\eta |}\, e^{i(X\cdot\eta + 2X'\cdot\eta ')}\, F_\alpha (\eta +\eta ')\, F_{\alpha '}^A(-\eta ; \eta ')\, \int_{\R^3}\, e^{-2i\eta '\cdot y}\, \tilde\chi (y)\\
 &\hspace{5.5cm}\times\, a_\alpha (y; \ux_j)\, \overline{a'}_{\alpha '}(y; \ux_k')\; dy\, d\eta\, d\eta '\period
\end{align*}
Similarly, we can find some $C_Q''>0$ such that, for $|\eta|\geq 1$ and $|\eta|\geq 2|\eta '|$, 
\[\bigl|F_\alpha ^R(\eta ; \eta ')F_{\alpha '}^A(-\eta ; \eta ')\bigr|\ \leq\ C_Q''\, |\eta|^{-8-|\alpha|-|\alpha '|-Q}\, |\eta '|^{2Q}\]
and obtain $C_{\alpha ; \alpha '}^A\sim C_{\alpha ; \alpha '}^{A_2}$ where 
\begin{align*}
 C_{\alpha ; \alpha '}^{A_2}(X; X'; \ux _j; \ux _k)\ =&\ \int_{|\eta|\geq 1\atop 2|\eta '|\leq |\eta |}\, e^{i(X\cdot\eta + 2X'\cdot\eta ')}\, F_\alpha ^A(\eta ; \eta ')\, F_{\alpha '}^A(-\eta ; \eta ')\, \int_{\R^3}\, e^{-2i\eta '\cdot y}\, \tilde\chi (y)\\
 &\hspace{5.5cm}\times\, a_\alpha (y; \ux_j)\, \overline{a'}_{\alpha '}(y; \ux_k')\; dy\, d\eta\, d\eta '\period
\end{align*}
As above, adding to this integral the contribution of the regions $\{(\eta ; \eta ');\, |\eta|\geq 1\comma\, |\eta|<2|\eta '|\}$ and 
$\{\eta ;\, |\eta|<1\}$ amounts to add to the function $C_{\alpha ; \alpha '}^{A_2}$ a smooth function. Therefore $C_{\alpha ; \alpha '}^A\sim C_{\alpha ; \alpha '}^B$ where 
\begin{align*}
 C_{\alpha ; \alpha '}^B(X; X'; \ux _j; \ux _k)\ =&\ \int_{\R^6}\, e^{i(X\cdot\eta + 2X'\cdot\eta ')}\, F_\alpha ^A(\eta ; \eta ')\, F_{\alpha '}^A(-\eta ; \eta ')\, \int_{\R^3}\, e^{-2i\eta '\cdot y}\, \tilde\chi (y)\\
 &\hspace{5.5cm}\times\, a_\alpha (y; \ux_j)\, \overline{a'}_{\alpha '}(y; \ux_k')\; dy\, d\eta\, d\eta '\period
\end{align*}
By the multinomial theorem, we have 
\[\frac{1}{q!}\, \bigl(\eta '\cdot \nabla\bigr)^qF_\alpha \ =\ \frac{1}{q!}\, \sum _{|\beta|=q}\, \frac{q!}{\beta !}\, (\eta ')^\beta\, \partial^\beta F_\alpha\ =\ \sum _{|\beta|=q}\, \frac{1}{\beta !}\, (i\eta ')^\beta F_{\alpha +\beta}\comma\]
since $F_\alpha=(-i\partial)^\alpha F_0$. This yields 
\begin{align*}
 C_{\alpha ; \alpha '}^B(X; X'; \ux _j; \ux _k)\ =&\ \sum _{|\beta|<Q\atop |\beta '|<Q}\, \frac{1}{\beta ! \beta '!}\, \int_{\R^3}\, e^{iX\cdot\eta}\, F_{\alpha +\beta}(\eta)\, F_{\alpha '+\beta '}(-\eta)\, d\eta\\
 &\hspace{-0.5cm}\times\, \int_{\R^3}\, e^{2iX'\cdot\eta '}\, (i\eta ')^{\beta +\beta '}\, \int_{\R^3}\, e^{-2i\eta '\cdot y}\, \tilde\chi (y)\, a_\alpha (y; \ux_j)\, \overline{a'}_{\alpha '}(y; \ux_k')\; dy\, d\eta '\period
 \end{align*}
Since 
\begin{align*}
 &\ \int_{\R^3}\, e^{2iX'\cdot\eta '}\, (i\eta ')^{\beta +\beta '}\, \int_{\R^3}\, e^{-2i\eta '\cdot y}\, \tilde\chi (y)\, a_\alpha (y; \ux_j)\, \overline{a'}_{\alpha '}(y; \ux_k')\; dy\, d\eta '\\
 =&\ (\partial _{X'}/2)^{\beta +\beta '}\, \int_{\R^3}\, e^{2iX'\cdot\eta '}\, \int_{\R^3}\, e^{-2i\eta '\cdot y}\, \tilde\chi (y)\, a_\alpha (y; \ux_j)\, \overline{a'}_{\alpha '}(y; \ux_k')\; dy\, d\eta '\\
 =&\ (2\pi)^3\, (\partial _y/2)^{\beta +\beta '}\, \bigl(\tilde\chi (y)\, a_\alpha (y; \ux_j)\, \overline{a'}_{\alpha '}(y; \ux_k')\bigr)_{|y=X'}
\end{align*}
by \eqref{eq:fourier-inv-g-bis}, we get 
\begin{align}
 C_{\alpha ; \alpha '}^B(X; X'; \ux _j; \ux _k)\ =&\ (2\pi)^3\, \sum _{|\beta|<Q\atop |\beta '|<Q}\, \frac{1}{\beta ! \beta '!}\, \int_{\R^3}\, e^{iX\cdot\eta}\, F_{\alpha +\beta}(\eta)\, F_{\alpha '+\beta '}(-\eta)\, d\eta\label{eq:dev-C-B}\\
 &\hspace{2.0cm}\times\, (\partial _y/2)^{\beta +\beta '}\, \bigl(\tilde\chi (y)\, a_\alpha (y; \ux_j)\, \overline{a'}_{\alpha '}(y; \ux_k')\bigr)_{|y=X'}\nonumber
 \period
\end{align}
Recall that we worked for $n_j\leq |\alpha|<m+4$ and $n_k'\leq |\alpha '|<m+4$ with $m>4+n_j+n_k'$. By the properties of the functions $F_\alpha$ (cf. \eqref{eq:comportement-F-alpha}) and Lemma~\ref{lm:inté-parties}, a term in the sum in \eqref{eq:dev-C-B} belongs to the class $\rC^{4+|\alpha |+|\beta|+|\alpha '|+|\beta '|}$. In particular, it belongs to the class $\rC^m$ if $4+|\alpha |+|\beta|+|\alpha '|+|\beta '|\geq m$. \\
Coming back to the inverse Fourier transform $F_{F^A}^i$ of $F^A$ in \eqref{eq:fourier-inv-F-A}, we can write 
\begin{align*}
&F_{F^A}^i\bigl(X'+X/2; X'-X/2; \ux _j; \ux _k\bigr)\ \sim\ G\bigl(X; X'; \ux _j; \ux _k\bigr)\hspace{.4cm}\mbox{where}\\
&\ G\bigl(X; X'; \ux _j; \ux _k\bigr)\\
\ =&\ (2\pi)^{-3}\, \chi _j\bigl(\ux_j\bigr)\; \chi _k'\bigl(\ux_k'\bigr)\; \sum _{n_j\leq |\alpha|,\, n_k'\leq |\alpha '|\atop 4+|\alpha|+|\alpha '|<m}\, \sum _{4+|\alpha|+|\beta|+|\alpha '|+|\beta '|<m}\, \frac{1}{\beta ! \beta '!}\\ 
&\hspace{5.0cm}\times\, \int_{\R^3}\, e^{iX\cdot\eta}\, F_{\alpha +\beta}(\eta)\, F_{\alpha '+\beta '}(-\eta)\, d\eta\\
&\hspace{5.0cm}\times\, (\partial _y/2)^{\beta +\beta '}\, \bigl(\tilde\chi (y)\, a_\alpha (y; \ux_j)\, \overline{a'}_{\alpha '}(y; \ux_k')\bigr)_{|y=X'}\period
\end{align*}
We showed at the beginning of the proof that $\tilde\gamma _0\sim F_{F^A}^i$. Thus 
\[\gamma _0\bigl(X; X'; \ux _j; \ux _k\bigr)\ \sim \ F_{F^A}^i\bigl(X'+X/2; X'-X/2; \ux _j; \ux _k\bigr)\period\]
Defining $\rR_m$ by $\gamma _0-G$, we obtain \eqref{eq:dev-gamma-1}. $\cqfd$

We now focus on the integrals remaining in \eqref{eq:dev-gamma-1}, using Lemma~\ref{lm:comportement-F-alpha} and \eqref{eq:dérivée-alpha-norme-4}. 
\begin{lemma}\label{lm:régu-fourier-inv}
Let $(\alpha ; \alpha ')\in (\N^3)^2$. There exists a smooth function $S:\, \R^3\dans\C$ such that, for all $X\in \R^3$, 
\begin{align}\label{eq:régu-fourier-inv}
\int_{\R^3}\, e^{iX\cdot\eta}\, F_\alpha(\eta)\, F_{\alpha '}(-\eta)\, d\eta\ =&\ \frac{-8\, (2\pi )^4}{(6+2|\alpha|+2|\alpha '|)!}\, \, \Bigl(P_\alpha (-\partial_x)\, P_{\alpha '} (\partial_x)\, |x|^{5+2|\alpha|+2|\alpha '|}\Bigr)_{|x=X}\nonumber\\
&\, +\, S(X)\period
\end{align}
\end{lemma}
\Pf For two functions $f$ and $g$, we write here $f\sim g$ if $f-g$ is a smooth function. We shall frequently use the following fact: if $f\in\rC_c^\infty (\R)$ and $g\in\rC^\infty (\R^3)$ then the map 
\begin{equation}\label{eq:supp-compact}
 \R^3\ni X\ \donne\ \int_{\R^3}\, e^{iX\cdot\eta}\, f\bigl(|\eta|\bigr)\, g(\eta)\, d\eta
\end{equation}
is smooth. This indeed follows from Lemma~\ref{lm:inté-parties}. Denote by $N(X)$ the l.h.s. of \eqref{eq:régu-fourier-inv}. \\
Consider a cut-off function $\tau\in\rC^\infty (\R^+; \R)$ such that $\tau =0$ on $[0; 1]$ and $\tau =1$ on $[2; +\infty[$. Since $(1-\tau)\in\rC_c^\infty (\R)$, we can write, by \eqref{eq:supp-compact}, that $N\sim N_1$ where 
\[N_1(X)\ =\ \int_{\R^3}\, e^{iX\cdot\eta}\, \tau \bigl(|\eta|\bigr)\, F_\alpha(\eta)\, F_{\alpha '}(-\eta)\, d\eta\period\]
We use the decomposition \eqref{eq:decomp-F-alpha} for $F_\alpha$, the estimate \eqref{eq:comportement-G-alpha}, and Lemma~\ref{lm:inté-parties}, to see that $N\sim N_2$ where 
\begin{align}
 N_2(X)\ =&\ (8\pi)^2\, \int_{\R^3}\, e^{iX\cdot\eta}\, \tau \bigl(|\eta|\bigr)\, \bigl((-i\partial^\alpha )|\cdot|^{-4}\bigr)(\eta)\, \bigl((-i\partial^{\alpha '})|\cdot|^{-4}\bigr)(-\eta)\, d\eta\nonumber\\
 =&\ (8\pi)^2\, i^{|\alpha|+|\alpha '|}\, \int_{\R^3}\, e^{iX\cdot\eta}\, \tau \bigl(|\eta|\bigr)\, P_\alpha (\eta)\, P_{\alpha '}(-\eta)\, |\eta|^{-8-2|\alpha|-2|\alpha '|}\, d\eta\period\label{eq:N2}
\end{align}
Since the function $\R^3\ni\eta\donne \tau (|\eta|)|\eta|^{-8}$ is integrable, standard derivation under the integral shows that 
\[N_2(X)\ =\ (8\pi)^2\, P_\alpha (-\partial _X)\, P_{\alpha '} (\partial _X)\, N_3(X)\period\]
where 
\[N_3(X)\ =\ \int_{\R^3}\, e^{iX\cdot\eta}\, \tau \bigl(|\eta|\bigr)\, |\eta|^{-8-2|\alpha|-2|\alpha '|}\, d\eta\period\]
Using spherical coordinates (see the appendix in \cite{jn}), we get, for $X\neq 0$, 
\[N_3(X)\ =\ 4\pi\, \int_1^{+\infty}\, \tau (r)\, r^{-7-2|\alpha|-2|\alpha '|}\; \frac{\sin (r|X|)}{|X|}\, dr\comma\]
since the support of $\tau$ is included in $[1; +\infty[$. By integration by parts, 
\begin{align*}
 N_3(X)\ =&\ \frac{4\pi}{-6-2|\alpha|-2|\alpha '|}\, \left(\left[\tau (r)\, \frac{\sin (r|X|)}{|X|}\, r^{-6-2|\alpha|-2|\alpha '|}\right]_{r=1}^{r=+\infty}\right.\\
 &\hspace{4.0cm}\left.\, -\, \int_1^{+\infty}\, r^{-6-2|\alpha|-2|\alpha '|}\; \partial_r\left(\tau (r)\, \frac{\sin (r|X|)}{|X|}\right)\, dr\right)\\
 =&\ \frac{4\pi}{6+2|\alpha|+2|\alpha '|}\, \int_1^{+\infty}\, r^{-6-2|\alpha|-2|\alpha '|}\; \partial_r\left(\tau (r)\, \frac{\sin (r|X|)}{|X|}\right)\, dr\comma
\end{align*}
since $\tau$ is flat at $1$. The contribution of the derivative of $\tau$ in the previous integral is a smooth function by \eqref{eq:supp-compact}. Thus 
\[N_3(X)\ \sim\ \frac{4\pi}{6+2|\alpha|+2|\alpha '|}\, \int_1^{+\infty}\, r^{-6-2|\alpha|-2|\alpha '|}\; \tau (r)\, \cos (r|X|)\, dr\period\]
Repeating the integration by parts, we get, by a finite induction, that 
\[N_3(X)\ \sim\ \frac{4\pi\, |X|^{4+2|\alpha|+2|\alpha '|}}{(6+2|\alpha|+2|\alpha '|)!}\, \int_1^{+\infty}\, r^{-2}\; \tau (r)\, \cos (r|X|)\, dr\period\]
Now, using again an integration by parts, we write 
\begin{align*}
 \int_1^{+\infty}\, r^{-2}\; \tau (r)\, \cos (r|X|)\, dr\ =&\ \lim_{R\to +\infty}\, \int_1^R\, r^{-2}\; \tau (r)\, \cos (r|X|)\, dr\\
 \ \sim&\ -\, \lim_{R\to +\infty}\, |X|\, \int_1^R\, r^{-1}\; \tau (r)\, \sin (r|X|)\, dr\\
 \ \sim&\ -\, |X|\, \lim_{R\to +\infty}\, \int_0^{R}\, r^{-1}\; \sin (r|X|)\, dr\\
 \ \sim&\ -\, |X|\, \lim_{R\to +\infty}\, \int_0^{R|X|}\, s^{-1}\; \sin (s)\, ds\\
 \ \sim&\ -\, \frac{\pi}{2}\, |X|\
\end{align*}
since the semi-convergent Dirichlet integral equals $\pi/2$. Thus 
\[N_2(X)\ \sim\ \frac{-8\, (2\pi )^4\, i^{|\alpha|+|\alpha '|}}{(6+2|\alpha|+2|\alpha '|)!}\, P_\alpha (-\partial _X)\, P_{\alpha '} (\partial _X)\, |X|^{5+2|\alpha|+2|\alpha '|}\comma\]
yielding \eqref{eq:régu-fourier-inv}. $\cqfd$

Inserting the result of Lemma~\ref{lm:régu-fourier-inv} into \eqref{eq:dev-gamma-1}, we get  
\begin{align}
&\ \gamma _0 \bigl(X; X';\, \ux_j;\, \ux _k'\bigr)\ -\ \tilde\rR_m\bigl(X; X';\, \ux_j;\, \ux _k'\bigr)\label{eq:dev-gamma-2}\\
\ =&\ \chi _j\bigl(\ux_j\bigr)\; \chi _k'\bigl(\ux_k'\bigr)\; \sum _{n_j\leq |\alpha |\, ,\  n_k'\leq |\alpha '|\atop 4+|\alpha|+|\alpha '|<m}\ \sum _{4+|\alpha|+|\beta|+|\alpha '|+|\beta '|<m}\, \frac{1}{\beta !\, \beta '!}\nonumber\\
&\hspace{4.5cm}\times
\Bigl(\frac{\partial _y}{2}\Bigr)^{\beta +\beta '}\Bigl(\tilde\chi (y)\, a_\alpha (y; \ux_j)\, \overline{a'}_{\alpha '}(y; \ux_k')\Bigr)_{|y=X'}\nonumber\\
&\hspace{4.5cm}\times\, \frac{-8\, (2\pi )^4}{(6+2|\alpha|+2|\alpha '|+2|\beta|+2|\beta '|)!}\nonumber\\
&\hspace{4.5cm}\times\, \Bigl(P_{\alpha +\beta}(-\partial_x)\, P_{\alpha '+\beta '} (\partial_x)\, |x|^{5+2|\alpha|+2|\alpha '|+2|\beta|+2|\beta '|}\Bigr)_{|x=X}\comma\nonumber
\end{align}

for some function $\tilde\rR_m : \cR\times\cV_{\neq}\times\cV_{\neq}'\dans\C$, that belongs to the class $\rC^m$. \\
In the expansion \eqref{eq:dev-gamma-2}, it is natural to expect that the regularity of $\gamma _0$ is the one of 
\begin{align}
&\ T\bigl(X; X';\, \ux_j;\, \ux _k'\bigr)\label{eq:T}\\
\ =&\ \frac{-16\pi }{(6+2n_j+2n_k')!}\; \chi _j\bigl(\ux_j\bigr)\; \chi _k'\bigl(\ux_k'\bigr)\; \tilde\chi (X')\; \sum _{|\alpha |\, =\, n_j\atop|\alpha '|\, =\, n_k'}\, a_\alpha (X'; \ux_j)\, \overline{a'}_{\alpha '}(X'; \ux_k')\nonumber\\
&\hspace{7.4cm}\times\, \Bigl(P_\alpha (-\partial_x)\, P_{\alpha '} (\partial_x)\, |x|^{5+2n_j+2n_k'}\Bigr)_{|x=X}\comma\nonumber
\end{align}
that is the sum of the terms, with indices $(\alpha ; \beta; \alpha'; \beta ')$ such that $|\alpha|=n_j$, $\beta=0$, $|\alpha '|=n_k'$, and $\beta '=0$, in the sum on the r.h.s. of \eqref{eq:dev-gamma-2}. That is precisely what Lemma~\ref{lm:non-nul} and Proposition~\ref{prop:régu-gamma} below prove. 
\begin{lemma}\label{lm:non-nul}
Let $T : (\R^3)^{2(N-1)}\dans\C$ be the function defined by \eqref{eq:T}. Then, for $(X; X';\, \ux_j;\, \ux _k')\in (\R^3)^{2(N-1)}$, 
\begin{align}
&\ T\bigl(X; X';\, \ux_j;\, \ux _k'\bigr)\label{eq:T-varphi}\\
\ =&\ \frac{-16\pi }{(6+2n_j+2n_k')!}\; \chi _0\bigl(X'; \ux_j\bigr)\; \chi _0'\bigl(X'; \ux_k'\bigr)\; \sum _{|\alpha |\, =\, n_j\atop|\alpha '|\, =\, n_k'}\, \varphi_\alpha (X'; \ux_j)\, \overline{\varphi '}_{\alpha '}(X'; \ux_k')\nonumber\\
&\hspace{7.5cm}\times\, \Bigl(P_\alpha (-\partial_x)\, P_{\alpha '} (\partial_x)\, |x|^{5+2n_j+2n_k'}\Bigr)_{|x=X}\comma\nonumber
\end{align}
where the functions $\varphi _\alpha$ appear in the formula \eqref{eq:dév-varphi-j} and the functions $\varphi _{\alpha '}'$ appear in the formula \eqref{eq:dév-varphi'-k}.
Moreover, the regularity of the function $T$ near $(0; \hat{x}_j; \hat{\ux}_j; \hat{\ux}_k')$ is $4+n_j+n_k'$. 
\end{lemma}
\Pf The equality \eqref{eq:T-varphi} immediately follows from Proposition~\ref{prop:dév-fourier-gamma} and the properties of the cut-off functions $\chi_0$, $\chi_0'$, $\chi_{0; j}$, $\chi_{0; k}'$, and $\tilde\chi$. By Proposition~\ref{prop:dév-gamma} and Lemma~\ref{lm:régu-fourier-inv}, the function $T$ belongs to the class $\rC^{4+n_j+n_k'}$. \\
Assume that $T$ belongs to the class $\rC^{5+n_j+n_k'}$. By the properties of $\chi_0$ and $\chi_0'$, there exist some non empty open sets $\cU$ of $\R^3$, $\cO$ of $(\R^3)^{(N-1)}$, and $\cO'$ of $(\R^3)^{(N-1)}$, such that, for all $(X';\, \ux_j;\, \ux _k')\in\cU\times\cO\times\cO'$, the product $\chi _0(X'; \ux_j)\chi _0'(X'; \ux_k')=1$. \\
Let $(X';\, \ux_j;\, \ux _k')\in\cU\times\cO\times\cO'$. We know from the properties of the functions $P_\alpha$ and \eqref{eq:T} 
that, on $\R^3\setminus\{0\}$, any partial derivative of the map $X\donne T(X; X'; \ux _j; \ux_k')$ of order $5+n_j+n_k'$ is homogeneous of degre zero and therefore a function of $X/|X|$ only. We assume that such a derivative exists at $0$ and is continuous there. Thus this derivative must be constant. In particular, this map is smooth on $\R^3$. Since $T$ is smooth w.r.t. the other variables, $T$ is smooth. \\
Let $\tau\in\rC_c^\infty(\R)$ such that $\tau =1$ near $0$. Since $\tau$ is flat at zero and the map $x\donne |x|^{5+2n_j+2n_k'}$ is smooth away from zero, the map $T_1: \R^3\times\cU\times\cO\times\cO'\dans\C$ given by 
\begin{align*}
&\ T_1\bigl(X; X';\, \ux_j;\, \ux _k'\bigr)\\ 
=&\ \sum _{|\alpha |\, =\, n_j\atop|\alpha '|\, =\, n_k'}\, \varphi_\alpha (X'; \ux_j)\, \overline{\varphi '}_{\alpha '}(X'; \ux_k')\, \Bigl(P_\alpha (-\partial_x)\, P_{\alpha '} (\partial_x)\, \tau \bigl(|x|\bigr)\, |x|^{5+2n_j+2n_k'}\Bigr)_{|x=X}
\end{align*}
is smooth as well. Let $K$ be a compact subset of $\cU\times\cO\times\cO'$.
By the proof of Lemma~\ref{lm:inté-parties}, there exists some $c>0$ such that, for all $(X';\, \ux_j;\, \ux _k')\in\cU\times\cO\times\cO'$, for all $\eta\in\R^3$ with $|\eta|\geq 1$, 
\begin{equation}\label{eq:borne-fourier-T1}
 \bigl|F_{T_1} \bigl(\eta; X';\, \ux_j;\, \ux _k'\bigr)\bigr|\ \leq\ c\, |\eta |^{-(9+n_j+n_k')}\period
\end{equation}
Let us denote by $f$ the Fourier transform of the map $x\donne \tau (|x|)|x|^{5+2n_j+2n_k'}$. Let $\cU_1:=\{\eta\in\R^3; |\eta|>1\}$. For $(\eta; X';\, \ux_j;\, \ux _k')\in\cU_1\times\cU\times\cO\times\cO'$, we have 
\[F_{T_1} \bigl(\eta; X';\, \ux_j;\, \ux _k'\bigr)\ =\ f(\eta)\, \sum _{|\alpha |\, =\, n_j\atop|\alpha '|\, =\, n_k'}\, 
\varphi_\alpha (X'; \ux_j)\, \overline{\varphi '}_{\alpha '}(X'; \ux_k')\, P_\alpha (-i\eta)\, P_{\alpha '}(i\eta)\period\]
By Lemma~\ref{lm:fourier-norme-localisée} applied to $f$, 
\[F_{T_1} \bigl(\eta; X';\, \ux_j;\, \ux _k'\bigr)\ =\ \frac{\lambda}{|\eta|^{8+2n_j+2n_k'}}\, \sum _{|\alpha |\, =\, n_j\atop|\alpha '|\, =\, n_k'}\, \varphi_\alpha (X'; \ux_j)\, \overline{\varphi '}_{\alpha '}(X'; \ux_k')\, P_\alpha (-i\eta)\, P_{\alpha '}(i\eta)\, +\, g(\eta)\]
and $|g(\eta)|\leq c|\eta|^{-(9+n_j+n_k')}$, for some $\lambda\neq 0$ and some $c>0$. 
By \eqref{eq:borne-fourier-T1} and homogeneity, the last double sum must be zero identically on 
$\cU_1\times\cU\times\cO\times\cO'$. Thus, the product 
\[\biggl(\sum _{|\alpha |\, =\, n_j}\, \varphi_\alpha (X'; \ux_j)\, P_\alpha (-i\eta)\biggr)\, \biggl(\sum _{|\alpha '|\, =\, n_k'}\, \overline{\varphi '}_{\alpha '}(X'; \ux_k')\, P_{\alpha '}(i\eta)\biggr)\]
is zero identically on $\cU_1\times\cU\times\cO\times\cO'$. Since it is a product of real analytic functions, one factor must be zero on $\cU_1\times\cU\times\cO\times\cO'$ (cf. \cite{ca}). By Lemma~\ref{lm:famille-libre}, this implies that either all the functions $\varphi _\alpha$ with $|\alpha|=n_j$ are zero on $\cU\times\cO$ or all the functions $\varphi _{\alpha '}'$ with $|\alpha '|=n_k'$ are zero on $\cU\times\cO'$. This contradicts the properties of these functions stated just after \eqref{eq:dév-varphi-j} and \eqref{eq:dév-varphi'-k}, respectively. \\
The regularity of $T$ near $(0; \hat{x}_j; \hat{\ux}_j; \hat{\ux}_k')$ is therefore $4+n_j+n_k'$. $\cqfd$

By Proposition~\ref{prop:dév-gamma}, we know that all the terms on the r.h.s. of \eqref{eq:dev-gamma-2} but T belong to the class $\rC^{5+|\alpha |+|\alpha '|}$. Thus \eqref{eq:dev-gamma-2} and Lemma~\ref{lm:non-nul} show that the regularity of $\gamma_0$ near $(0; \hat{x}_j; \hat{\ux}_j; \hat{\ux}_k')$ is precisely the one of $T$, that is $4+n_j+n_k'$. We have proven 
\begin{proposition}\label{prop:régu-gamma}
The regularity of $\gamma _0$ near $(0; \hat{x}_j; \hat{\ux}_j; \hat{\ux}_k')$ is $4+n_j+n_k'$. So is also the one of $\tilde\gamma _0$ near $(\hat{x}_j; \hat{x}_j; \hat{\ux}_j; \hat{\ux}_k')$. 
\end{proposition}

Now, we are able to prove our main result. 

\Pfof{Theorem~\ref{th:régu-matice-densité}} Let us take a neighbourhood $\cN$ of $\hat{x}$ and a neighbourhood $\cN'$ of $\hat{x}'$ such that $\chi _0=1$ on $\cN$ and $\chi _0'=1$ on $\cN'$. Thanks to the Propositions~\ref{prop:mod-C-infini} and \ref{prop:dév-gamma} and to the Lemmata~\ref{lm:régu-fourier-inv} and \ref{lm:non-nul}, we get \eqref{eq:dév-gamma-N-1} and \eqref{eq:formule-T-j}. Proposition~\ref{prop:régu-gamma} provides the regularity of each $T_j$ and \eqref{eq:formule-T-j} shows that the less regular term in $T_j$ cannot be compensated by a term coming from some $T_\ell$ with $\ell\neq j$. Therefore, the regularity of $\gamma_{N-1}$ is the minimum of the set $\{4+n_j+n_{c(j)}';\, j\in\cD\}$ of regularities, that is $4+p$. $\cqfd$

\section{Pseudodifferential structure of operators associated to localisations of the density matrix.}
\label{s:pseudo}
\setcounter{equation}{0}

In this section, we consider the integral operator $\Gamma$, the kernel of which is the density matrix $\gamma _{N-1}$. It naturally acts on squared integrable functions $f$ on $\R^{3(N-1)}$ as 
\[\bigl(\Gamma f\bigr)(\ux)\ =\ \int_{\R^{3(N-1)}}\, \gamma _{N-1}(\ux; \ux')\, f(\ux')\, d\ux'\]
and is actually a bounded, self-adjoint operator on $\rL^2(\R^{3(N-1)})$. We focus on localisations of this operator of the form $\Gamma^0=\chi _0\Gamma\chi _0'$ for cut-off functions $\chi_0\in\rC_c^\infty(\R^{3(N-1)}; \R)$ and $\chi_0'\in\rC_c^\infty(\R^{3(N-1)}; \R)$. Precisely, $\Gamma^0$ is the composition of the multiplication operator by $\chi _0'$, of $\Gamma$, and of the multiplication operator by $\chi _0$, in this order. \\
If the tensor product $\chi _0\otimes\chi _0'$ localises on a region where $\gamma _{N-1}$ is smooth then $\Gamma^0$ transforms $\rL^2(\R^{3(N-1)})$-functions to a smooth function. Here we are more interested in the case where this tensor product localises near a point 
\[(\hat{\ux}; \hat{\ux}')\in \bigl(\cU^{(1)}_{N-1}\times\cU^{(1)}_{N-1}\bigr)\, \cap\, \cC _{N-1}^{(2)}\]
as in Section~\ref{s:Fourier-matrice}. If $\hat{\ux}=\hat{\ux}'$, it was shown in \cite{jn} (Section 5) that $\Gamma^0$ can be viewed as a pseudodifferential operator, the symbol of which belongs to a well-known class of smooth symbols. This property was detected after a study of the wave front set of $\gamma _{N-1}$. In the same spirit, we now use Theorem~\ref{th:régu-matice-densité} and the notion of singular support (cf. \cite{h2}, p. 42) to detect a regular pseudodifferential structure in $\Gamma^0$. First of all, we introduce the notion of global, smooth symbols that is studied in \cite{h3}, Chapter 18.1. \\
For positive intergers $n$ and $p$, for $m\in\R$, let $\rrS^m(\R^n\times\R^p)$ be the set of smooth functions $a$ on $\R^n\times\R^p$ such that, for all $(\alpha ; \beta)\in (\N^n\times\N^p)$, 
\[\sup _{(\ux; \uxi)\in\R^n\times\R^p}\, \langle \uxi\rangle^{-m+|\beta|}\, \Bigl|\bigl(\partial_\ux^\alpha\partial_\uxi^\beta a\bigr)\bigl(\ux ; \uxi\bigr)\Bigr|\ <\ +\infty\period\]
We denote by $\rrS^{-\infty}$ the intersection of all such space $\rrS^m$. \\
It is well-known (see Theorem 18.1.6. in \cite{h3}) that, if $K$ is the kernel of a pseudodifferential operator on $\R^{3(N-1)}$, the symbol of which belongs to the class $\rrS^m(\R^{3(N-1)}\times\R^{3(N-1)})$, for some $m\in\R$, then its singular support must be contained in the diagonal of $(\R^{3(N-1)})^2$, namely 
\[D\ :=\ \Bigl\{(\ux; \ux)\in(\R^{3(N-1)})^2 \, ;\ \ux\in\R^{3(N-1)}\, \Bigr\}\period\]
According to \cite{h3}, Chapter 18.1, $\Gamma^0$ can always be considered as a pseudodifferential operator. 
But, from Theorem~\ref{th:régu-matice-densité}, we see that the singular support of $\gamma _{N-1}$ inside $\cN\times\cN'$ contains points $(\ux; \ux')$ such that, for all $j\in\cD$, $x_j=x_{c(j)}'$. These points do not belong to $D$ unless the map $c$ is the identity on $\cD$. \\
These considerations suggest, in the general case, that an appropriate permutation of some variables should reveal a regular pseudodifferential structure in $\Gamma^0$. That is precisely what we are going to show. 
Let us take a point 
\[(\hat{\ux}; \hat{\ux}')\in \bigl(\cU^{(1)}_{N-1}\times\cU^{(1)}_{N-1}\bigr)\, \cap\, \cC _{N-1}^{(2)}\]
and consider the map $c : \cD\dans \ncg 1; N-1\ncd$, that is associated to it. Since $c$ is injective, the sets $c(\cD)\setminus\cD$ and $\cD\setminus c(\cD)$ have the same, finite cardinal (possibly $0$). We choose an arbitrary bijection $b : c(\cD)\setminus\cD\dans \cD\setminus c(\cD)$. Now, we define a permutation $\sigma$ of $\ncg 1; N-1\ncd$ in the following way: for $\ell\in \ncg 1; N-1\ncd$, we set 
\[\sigma (\ell)=\ell\comma\ \mbox{if}\ \ell\not\in (\cD\, \cup\, c(\cD))\, ;\hspace{.4cm}\sigma (\ell)=b(\ell)\comma\ \mbox{if}\ \ell\in c(\cD)\setminus\cD\, ;\hspace{.4cm}\sigma (\ell)=c(\ell)\ \mbox{if}\ \ell\in \cD\period\]
We let $\sigma$ act on $\R^{3(N-1)}$ as 
\[\sigma\cdot\ux\ :=\ (x_{\sigma (\ell)})_{\ell\in\ncg 1; N-1\ncd}\ :=\ \bigl(x_{\sigma(1)};\, \cdots ; \, x_{\sigma(N-1)}\bigr)\period\]
It is a linear map and the modulus of its Jacobian is $1$. Its inverse $\sigma^{-1}\cdot$ acts on $\ux$ as  
$\sigma^{-1}\cdot\ux:=(x_{\sigma^{-1} (\ell)})_{\ell\in\ncg 1; N-1\ncd}$, where $\sigma^{-1}$ is the inverse of the permutation $\sigma$. We observe that the maps $\sigma\cdot$ and $\sigma^{-1}\cdot$ both preserve the set $\cU^{(1)}_{N-1}$. Let us define an unitary map $\rU : \rL^2(\R^{3(N-1)})\dans\rL^2(\R^{3(N-1)})$ by, for $f\in\rL^2(\R^{3(N-1)})$, $(\rU f)(\ux):=f(\sigma^{-1}\cdot \ux)$. Now, let us take cut-off functions $\chi _0$ and $\chi _0'$ as in Section~\ref{s:Fourier-matrice}. 
One can check that the kernel of $\Gamma^0\rU$ is given by 
\begin{equation}\label{eq:noyau-U}
 K(\ux ; \ux '')\ =\ \chi_0(\ux)\, \int_{\R^3}\, \psi (\ux ; y)\, \overline{\psi}(\sigma\cdot\ux ''; y)\, dy\ \chi _0''(\ux '')\comma
\end{equation}
if we define the cut-off function $\chi _0''$ by $\chi _0''(\ux '')=\chi _0'(\sigma\cdot\ux '')$. It  localises near the point $\hat{\ux}'':=\sigma^{-1}\cdot\hat{\ux}'$. By the choice of $\sigma$, the map $c$ associated to $(\hat{\ux}; \hat{\ux}'')$ is the identity on $\cD$. \\
As we shall see, the composition $\Gamma^0\rU$ is a pseudodifferential operator, the symbol of which belongs to some $\rrS^m$. According to Theorem 18.5.10 in \cite{h3}, we may choose the quantisation to see this property. It will be convenient to take the Weyl quantisation. If $L\in\rC_c^k((\R^{3(N-1)})^2; \C)$, its Weyl symbol $s_L$ is given by, for $(\ux ; \uxi)\in (\R^{3(N-1)})^2$,  
\begin{align}
&\ s_L(\ux ; \uxi)\nonumber\\
=&\ \int_{\R^{3(N-1)}}\, e^{-i\uxi\cdot\ut}\, L\bigl(\ux -\ut/2;\, \ux +\ut /2\bigr)\, d\ut\ =\ \int_{\R^{3(N-1)}}\, e^{i\uxi\cdot\ut}\, L\bigl(\ux +\ut/2;\, \ux -\ut /2\bigr)\, d\ut\comma\label{eq:def-symbole-weyl}
\end{align}
by the change of variables $\ut'=-\ut$. 

\begin{proposition}\label{prop:symbole-weyl}
Let $p$ be defined in \eqref{eq:p}. Then the Weyl symbol of the kernel $K$ belongs to the symbol class $\rrS^{-8-p}(\R^{3(N-1)}\times\R^{3(N-1)})$. Moreover, if a real $q>8+p$, this symbol does not belong to $\rrS^{-q}(\R^{3(N-1)}\times\R^{3(N-1)})$.
\end{proposition}
\begin{remark}\label{r:dev-asympt-symboles}
 Our proof of Proposition~\ref{prop:symbole-weyl} below provides a sequence of symbols such that the Weyl symbol of $K$ is the asymptotic sum of this sequence, in the sense of Proposition 18.1.3 in {\rm \cite{h3}}. \\
 We observe that $U$ acts on the bosonic {\rm(}resp. fermionic{\rm)} subspace of $\rL^2(\R^{3(N-1)})$ as a multiple of the identity operator. Therefore, by Proposition~\ref{prop:symbole-weyl}, the localised version $\Gamma^0$ of $\Gamma$ is a pseudodifferential operator with smooth symbol. 
\end{remark}

\Pfof{Proposition~\ref{prop:symbole-weyl}} First of all, we check that the treatment of $\gamma _{N-1}$ that was performed in Proposition~\ref{prop:mod-C-infini}, Lemma~\ref{lm:change-var}, Proposition~\ref{prop:dév-fourier-gamma}, Proposition~\ref{prop:dév-gamma}, Lemma~\ref{lm:régu-fourier-inv}, and Lemma~\ref{lm:non-nul}, can be applied to the kernel $K$. \\
In \cite{fhhs5} (see also Theorem 3.1 in \cite{jn}), a decomposition of $\psi$ near a two-particle collision is provided. Using the change of variables $\ux '=\sigma\cdot\ux ''$, such decomposition holds true for the map $(\ux ''; y)\mapsto\psi (\sigma \cdot\ux ''; y)$. 
This allows us to follow the proof of Proposition~\ref{prop:mod-C-infini} in \cite{jn} to get the statement of Proposition~\ref{prop:mod-C-infini} with $\gamma _{N-1}$ replaced by $K$, $\hat{\ux}'$ replaced by $\hat{\ux}''$, $\cV'$ replaced by some neighbourhood $\cV''$ of $\hat{\ux}''$, and, for $j\in\cD$, $c(j)$ replaced by $j$, and $\tilde\varphi _{c(j)}'$ replaced by the sum of a power series $\tilde\varphi _{j}''$ in the variables $(\ux ''; y)$. For $j\in\cD$, we have 
\[\tilde\varphi _{j}''(\ux ''; y)\ =\ \tilde\varphi _{c(j)}'(\sigma\cdot\ux ''; y)\period\]
In particular, the valuation of $\tilde\varphi _{j}''$ w.r.t. the variable $x_j''$ is exactly the one of $\tilde\varphi _{c(j)}'$ w.r.t. the variable $x_j'$. We thus have on $\cV\times\cV''$, for some smooth function $s$, 
\begin{equation}\label{eq:décomp-K}
K(\ux; \ux'')\ =\ s(\ux; \ux'')\, +\, \sum _{j\in\cD}\, K_j(\ux; \ux'')\comma\hspace{.4cm}\mbox{where}
\end{equation}
\begin{equation}\label{eq:K-j}
 K_j(\ux; \ux'')\ =\ \chi _0(\ux)\ \int_{\R^3}\, |x_j\, -\, y|\;  \tilde\varphi_j(\ux ; y) \; |x_j''\, -\, y|\; \overline{\tilde\varphi ''} _{j}(\ux ''; y)\; \tilde\chi _j(y)\, dy\ \chi _0''(\ux'')\comma
\end{equation}
 for $j\in\cD$. Recall that $\chi _0''(\ux)=\chi _0'(\sigma\cdot \ux)$. Let $j\in\cD$. Set 
\[\chi _j''\bigl(\ux_j\bigr)\ :=\ \prod_{\ell\in\ncg 1; N-1\ncd\atop \ell\neq c(j)}\, \chi _{0; \ell}'\bigl(x_{\sigma (\ell)}\bigr)\ =\ \prod_{q\in\ncg 1; N-1\ncd\atop q\neq j}\, \chi _{0; \sigma^{-1}(q)}'(x_q)\]
if $N>2$, else $\chi _j''=1$. In particular, $\chi _0''(\ux)=\chi _{0;c(j)}'(x_j)\, \chi_j''(\ux_j)$. \\
Since $K_j$ has the same structure as $(\chi _0\otimes\chi _0')\gamma _{N-1}^{(j)}$, the results of Proposition~\ref{prop:dév-fourier-gamma}, Proposition~\ref{prop:dév-gamma}, Lemma~\ref{lm:régu-fourier-inv}, and Lemma~\ref{lm:non-nul}, hold true for $K_j$ in place of $\tilde\gamma$ with the following changes: $k=c(j)$ is replaced by $k=j$, $n_k'$ replaced by $n_{c(j)}'$, $\chi _0'$ by $\chi _0''$, $\chi _k'$ by $\chi _j''$, $\chi _{0;k}'$ by $\chi _{0;c(j)}'$, $\cV_{\neq k}'$ by $\cV_{\neq j}''$, $\varphi _\alpha '$ by $\varphi _\alpha ''$ and $a_\alpha '$ by $a_\alpha ''$, where $\varphi _\alpha ''(\ux)=\varphi _\alpha '(\sigma\cdot\ux)$, and $a_\alpha ''(\ux)=a_\alpha '(\sigma\cdot\ux)$.
We denote by $\cR_j$ the range of the map $\cW_j^2\ni(x; y)\donne (x-y; (x+y)/2)$. It is a neighbourhood of $(0; \hat{x}_j)=(0; \hat{x}_j'')$. Then, for any integer $m$ larger than $4+n_j+n_{c(j)}'$, there exists a function $\rR_m^{(j)}\, :\, \cR_j\times\cV_{\neq j}\times\cV_{\neq j}''\dans\C$, that belongs to the class $\rC^m$, such that, for $(X; X'; \ux _j; \ux _j'')\in \cR_j\times\cV_{\neq j}\times\cV_{\neq j}''$, 
\begin{align}
&\ K_j\bigl(X/2+X'; X'-X/2;\, \ux_j;\, \ux _j''\bigr)\ -\ \rR_m^{(j)}\bigl(X; X';\, \ux_j;\, \ux _j''\bigr)\label{eq:dev-K_j}\\
\ =&\ \chi _j\bigl(\ux_j\bigr)\; \chi _j''\bigl(\ux_j''\bigr)\; \sum _{n_j\leq |\alpha |\, ,\  n_{c(j)}'\leq |\alpha '|\atop 4+|\alpha|+|\alpha '|<m}\ \sum _{4+|\alpha|+|\beta|+|\alpha '|+|\beta '|<m}\, \frac{1}{\beta !\, \beta '!}\nonumber\\
&\hspace{4.5cm}\times
\Bigl(\frac{\partial _y}{2}\Bigr)^{\beta +\beta '}\Bigl(\tilde\chi _j(y)\, a_\alpha (y; \ux_j)\, \overline{a''}_{\alpha '}(y; \ux_j'')\Bigr)_{|y=X'}\nonumber\\
&\hspace{4.5cm}\times\, \frac{-2\pi}{(6+2|\alpha|+2|\beta|+2|\alpha '|+2|\beta '|)!}\nonumber\\
&\hspace{4.5cm}\times\, \Bigl(P_{\alpha +\beta}(-\partial_x)\, P_{\alpha '+\beta '} (\partial_x)\, |x|^{5+2|\alpha|+2|\beta|+2|\alpha '|+2|\beta '|}\Bigr)_{|x=X}\period\nonumber
\end{align}
According to \eqref{eq:def-symbole-weyl}, the Weyl symbol $S_j$ of the operator with kernel $K_j$ is given by, for $(\ux ; \uxi)\in (\R^{3(N-1)})^2$,  
\[S_j(\ux ; \uxi)\ =\ \int_{\R^{3(N-1)}}\, e^{i\uxi\cdot\ut}\, K_j\bigl(x_j+t_j/2; x_j-t_j/2;\, \ux_j+\ut_j/2;\, \ux _j-\ut_j/2\bigr)\, d\ut\period\]
Using \eqref{eq:K-j} and a change of variables in the $y$-integral, we can show that $S_j$ is smooth (see the proof of Proposition 4.10 in \cite{jn} for details). By the localisation properties of $\chi _0$ and $\chi _0''$, $S_j$ vanishes outside a compact set in the variable $\ux$. \\
Let us consider the term with indices $(\alpha ; \beta; \alpha ' ; \beta ')$ and without the numerical factors in the double sum in \eqref{eq:dev-K_j}. Its Weyl symbol is given by 
\begin{align*}
&S_{\alpha ;\beta ; \alpha '; \beta '}^{(j)}(\ux ; \uxi)\nonumber\\
=&\int_{\R^{3(N-1)}}\, e^{i\uxi\cdot\ut}\, \chi _{0;j}\bigl(x_j+t_j/2\bigr)\, \chi _{0;c(j)}'\bigl(x_j-t_j/2\bigr)\, \chi _j\bigl(\ux_j+\ut_j/2\bigr)\; \chi _j''\bigl(\ux_j-\ut_j/2\bigr)\; \nonumber\\
&\hspace{1.5cm}\times\Bigl(\frac{\partial _y}{2}\Bigr)^{\beta +\beta '}\Bigl(\tilde\chi _j(y)\, a_\alpha \bigl(y; \ux_j+\ut_j/2\bigr)\, \overline{a''}_{\alpha '}\bigl(y; \ux_j-\ut_j/2\bigr)\Bigr)_{|y=x_j}\nonumber\\
&\hspace{1.5cm}\times\, \Bigl(P_{\alpha +\beta}(-\partial_x)\, P_{\alpha '+\beta '} (\partial_x)\, |x|^{5+2|\alpha|+2|\beta|+2|\alpha '|+2|\beta '|}\Bigr)_{|x=t_j}\, d\ut\period
\end{align*}
Let $\tilde\chi _{0;j}\in\rC_c^\infty(\R^3; \R)$ such that $\tilde\chi _{0;j}=1$ on the support of $\chi _{0;j}$ and $\tilde\chi _{0;j}\tilde\chi _j=\tilde\chi _{0;j}$. Let $\tilde\chi _{0;c(j)}'\in\rC_c^\infty(\R^3; \R)$ such that $\tilde\chi _{0;c(j)}'=1$ on the support of $\chi _{0;c(j)}'$ and $\tilde\chi _{0;c(j)}'\tilde\chi _j=\tilde\chi _{0;c(j)}'$. By Fubini's Theorem, the previous symbol splits into a product $S_{\alpha ;\beta ; \alpha '; \beta '}^{(j; \infty)}(\ux ; \uxi _j)S_{\alpha ;\beta ; \alpha '; \beta '}^{(j; j)}(x_j; \xi _j)$, where 
\begin{align*}
S_{\alpha ;\beta ; \alpha '; \beta '}^{(j; \infty)}(\ux ; \uxi _j)\ 
=&\ \int_{\R^{3(N-2)}}\, e^{i\uxi _j\cdot\ut_j}\, \chi _j\bigl(\ux_j+\ut_j/2\bigr)\; \chi _j''\bigl(\ux_j-\ut_j/2\bigr)\; \nonumber\\
&\hspace{1.2cm}\times\, \Bigl(\frac{\partial _y}{2}\Bigr)^{\beta +\beta '}\Bigl(\tilde\chi _j(y)\, a_\alpha \bigl(y; \ux_j+\ut_j/2\bigr)\, \overline{a''}_{\alpha '}\bigl(y; \ux_j-\ut_j/2\bigr)\Bigr)_{|y=x_j}\, d\ut_j
\end{align*}
and 
\begin{align*}
S_{\alpha ;\beta ; \alpha '; \beta '}^{(j; j)}(x_j; \xi _j)\ 
=&\ \int_{\R^3}\, e^{i\xi_j\cdot t_j}\, \tilde\chi _{0;j}\bigl(x_j+t_j/2\bigr)\, \tilde\chi _{0;c(j)}'\bigl(x_j-t_j/2\bigr)\nonumber\\
&\hspace{1.5cm}\times\ \Bigl(P_{\alpha +\beta}(-\partial_x)\, P_{\alpha '+\beta '} (\partial_x)\, |x|^{5+2|\alpha|+2|\beta|+2|\alpha '|+2|\beta '|}\Bigr)_{|x=t_j}\, dt_j\comma
\end{align*}
In both integrals, the integrand vanishes outside a compact set. By standart derivation under the integral sign, we see that $S_{\alpha ;\beta ; \alpha '; \beta '}^{(j; \infty)}$ (resp. $S_{\alpha ;\beta ; \alpha '; \beta '}^{(j; j)}$) is a smooth function of $(x_j; \ux_j; \uxi _j)$ (resp. $(x_j; \xi_j)$). By \eqref{eq:dev-K_j}, the Weyl symbol (see \eqref{eq:def-symbole-weyl}) of the map 
\begin{equation}\label{eq:reste-j}
 (\ux; \ux '')\ \donne\ \chi_0(\ux)\, \rR_m^{(j)}\bigl(x_j-x_j''; (x_j+x_j'')/2; \ux_j; \ux _j''\bigr)\, \chi_0''(\ux '')
\end{equation}
is also a smooth function. Since the function $\rR_m^{(j)}$ belongs to the class $\rC^m$, we can show, using $m$ times the identity
\eqref{eq:dérivée-exp} with $(x; \xi)$ replaced by $(\ut; \uxi)$, for nonzero $\uxi$, and integrations by parts, that the Weyl symbol of \eqref{eq:reste-j} belongs to $\rrS^{-m}(\R^{3(N-1)}\times\R^{3(N-1)})$.\\
We further observe that the function $S_{\alpha ;\beta ; \alpha '; \beta '}^{(j; \infty)}$ vanishes outside a compact set in the variable $\ux$ and so does $S_{\alpha ;\beta ; \alpha '; \beta '}^{(j; j)}$ outside a compact set in the variable $x_j$. \\ 
Using repetitively the identity \eqref{eq:dérivée-exp}, with $(x; \xi)$ replaced by $(\uxi _j; \ut _j)$ with nonzero $\uxi _j$, and integrations by parts, we see that 
$S_{\alpha ;\beta ; \alpha '; \beta '}^{(j; \infty)}\in \rrS^{-\infty}(\R^{3(N-1)}\times\R^{3(N-2)})$. \\
We now claim that $S_{\alpha ;\beta ; \alpha '; \beta '}^{(j; j)}\in\rrS^{-(8+|\alpha|+|\beta|+|\alpha '|+|\beta '|)}(\R^3\times\R^3)$. This property will imply that the symbol $S_{\alpha ;\beta ; \alpha '; \beta '}^{(j)}$ belongs to the class $\rrS^{-(8+|\alpha|+|\beta|+|\alpha '|+|\beta '|)}(\R^{3(N-1)}\times\R^{3(N-1)})$. \\
We first use \eqref{eq:régu-fourier-inv} in Lemma~\ref{lm:régu-fourier-inv} to write, for some $(\alpha ; \beta; \alpha'; \beta ')$-dependent constant $c$, 
\begin{align*}
S_{\alpha ;\beta ; \alpha '; \beta '}^{(j; j)}(x_j; \xi _j)\ 
=&\ c\int_{\R^3}\, e^{i\xi_j\cdot t_j}\, \tilde\chi _{0;j}\bigl(x_j+t_j/2\bigr)\, \tilde\chi _{0;c(j)}'\bigl(x_j-t_j/2\bigr)\\
&\hspace{1.5cm}\times\ \int_{\R^3}\, e^{it_j\cdot\eta}\, F_{\alpha +\beta}(\eta)\, F_{\alpha '+\beta '}(-\eta)\, d\eta\ \, dt_j\\
&\ +\ c\int_{\R^3}\, e^{i\xi_j\cdot t_j}\, \tilde\chi _{0;j}\bigl(x_j+t_j/2\bigr)\, \tilde\chi _{0;c(j)}'\bigl(x_j-t_j/2\bigr)\, S(t_j)\, dt_j\period
\end{align*}
Since the function $S$ is smooth, we can check, using the identity \eqref{eq:dérivée-exp}, with $(x; \xi)$ replaced by $(\xi _j; t_j)$ and integrations by parts, that the last integral belongs to $\rrS^{-\infty}(\R^3\times\R^3)$. By Fubini's theorem, we rewrite the previous double integral as 
\begin{align}
&c\int_{\R^3}\, F_{\alpha +\beta}(\eta)\, F_{\alpha '+\beta '}(-\eta)\nonumber\\
&\hspace{1.0cm}\times\ \int_{\R^3}\, 
e^{i(\xi_j+\eta)\cdot t_j}\, \tilde\chi _{0;j}\bigl(x_j+t_j/2\bigr)\, \tilde\chi _{0;c(j)}'\bigl(x_j-t_j/2\bigr)\, dt_j\ d\eta\nonumber\\
=&\ c\int_{\R^3}\, F_{\alpha +\beta}(\eta '-\xi_j)\, F_{\alpha '+\beta '}(\xi_j-\eta ')\, s_j(\eta ')\, d\eta '\comma\label{eq:int-eta'}\\
\mbox{where}&\hspace{1.0cm}s_j(\eta ')\ :=\ \int_{\R^3}\, 
e^{i\eta '\cdot t_j}\, \tilde\chi _{0;j}\bigl(x_j+t_j/2\bigr)\, \tilde\chi _{0;c(j)}'\bigl(x_j-t_j/2\bigr)\, dt_j\period\nonumber
\end{align}
Using again \eqref{eq:dérivée-exp}, with $(x; \xi)$ replaced by $(\eta '; t_j)$ and integrations by parts, $s_j$ satisfies 
\begin{equation}\label{eq:borne-poly}
 \forall\, k\in\N\comma\ \exists\, c_k>0\, ;\ \forall\, \eta'\in\R^3\setminus\{0\}\comma\ |\eta '|^k\, |s_j(\eta ')|\, \leq\, c_k\period
\end{equation}
Take $\xi_j\in\R^3$ with $|\xi_j|\geq 1$. By \eqref{eq:borne-poly} and the boundedness of the functions $F_\gamma$ (cf. \eqref{eq:comportement-F-alpha}), the contribution to the integral \eqref{eq:int-eta'} of the region $\{\eta '\in\R^3;\, |\eta '|\geq |\xi _j|/2\}$ is bounded above, for all $k\in\N$, by $2^kc_k|\xi _j|^{-k}$ times the $\rL^\infty$-norm of $F_{\alpha+\beta}$ times the one of $F_{\alpha'+\beta'}$. For the contribution of the complement, we use the Taylor expansions with exact integral remainder 
\[F_{\alpha +\beta}(\eta -\xi_j)\ =\ F_{\alpha +\beta}(-\xi_j)\, +\, \int_0^1\, \nabla F_{\alpha +\beta}(r\eta -\xi_j)\cdot\eta\, dr\]
and 
\[F_{\alpha '+\beta'}(\xi_j-\eta)\ =\ F_{\alpha '+\beta'}(\xi_j)\, -\, \int_0^1\, \nabla F_{\alpha '+\beta'}(\xi_j-r\eta)\cdot\eta\, dr\]
to get 
\begin{align*}
&\int_{|\eta |\leq |\xi _j|/2}\, F_{\alpha +\beta}(\eta -\xi_j)\, F_{\alpha '+\beta '}(\xi_j-\eta )\, s_j(\eta )\, d\eta \\
=&\ F_{\alpha +\beta}(-\xi_j)\, F_{\alpha '+\beta'}(\xi_j)\, \int_{|\eta '|\leq |\xi _j|/2}\, s_j(\eta )\, d\eta\ +\ R(\xi_j)
\end{align*}
such that $|R(\xi_j)|\leq d|\xi_j|^{-9-(|\alpha|+|\beta|+|\alpha '|+|\beta '|)}$, for some $\xi_j$-independent constant $d>0$, thanks to \eqref{eq:comportement-F-alpha}. Thus 
\[\int_{|\eta |\leq |\xi _j|/2}\, F_{\alpha +\beta}(\eta -\xi_j)\, F_{\alpha '+\beta '}(\xi_j-\eta )\, s_j(\eta )\, d\eta \ =\ F_{\alpha +\beta}(-\xi_j)\, F_{\alpha '+\beta'}(\xi_j)\, \int_{\R^3}\, s_j(\eta )\, d\eta\ +\ R'(\xi_j)\]
where $R'$ satisfies the same estimate as $R$, thanks to \eqref{eq:borne-poly}. Using \eqref{eq:comportement-F-alpha} again, we obtain that, uniformly w.r.t. $x_j$, 
\[S_{\alpha ;\beta ; \alpha '; \beta '}^{(j; j)}(x_j; \xi _j)\ =\ c\, |\xi_j|^{-8-(|\alpha|+|\beta|+|\alpha '|+|\beta '|)}\, \int_{\R^3}\, s_j(\eta )\, d\eta\, +\, O\bigl(|\xi_j|^{-9-(|\alpha|+|\beta|+|\alpha '|+|\beta '|)}\bigr)\period\]
Now, we observe that we may apply the above argument to any partial derivative of the function $S_{\alpha ;\beta ; \alpha '; \beta '}^{(j; j)}$. This proves the claim. \\
Coming back to $S_j$, we use \eqref{eq:dev-K_j} and the previous results for an integer $m>8+n_j+n_{c(j)}'$ to get $S_j=S_j^0+R$ where $R\in\rrS^{-9-n_j-n_{c(j)}'}(\R^{3(N-1)}\times\R^{3(N-1)})$ and $S_j^0$ is given, up to a multiplicative nonzero constant, by 
\begin{align*}
S_j^0(\ux ; \uxi)\ 
=&\ \sum _{|\alpha|=n_j,\atop |\alpha'|=n_{c(j)}'}\, \tilde\chi_j(x_j)\, \tilde\chi _{0;j}(x_j)\, \tilde\chi _{0;c(j)}'(x_j)\, F_\alpha(-\xi _j)\, F_{\alpha '}(\xi _j)\\
&\hspace{1.5cm}\times\ \int_{\R^{3(N-2)}}\, e^{i\uxi _j\cdot\ut_j}\, \chi _j\bigl(\ux_j+\ut_j/2\bigr)\; \chi _j''\bigl(\ux_j-\ut_j/2\bigr)\\
&\hspace{3.5cm}\times\ a_\alpha \bigl(x_j; \ux_j+\ut_j/2\bigr)\, \overline{a''}_{\alpha '}\bigl(x_j; \ux_j-\ut_j/2\bigr)\; d\ut_j\\
=&\ \tilde\chi _{0;j}(x_j)\, \tilde\chi _{0;c(j)}'(x_j)\, \int_{\R^{3(N-2)}}\, e^{i\uxi _j\cdot\ut_j}\, \chi _j\bigl(\ux_j+\ut_j/2\bigr)\; \chi _j''\bigl(\ux_j-\ut_j/2\bigr)\\
&\hspace{3.7cm}\times\ \sum_{|\alpha|=n_j}\, F_\alpha(-\xi _j)\, a_\alpha \bigl(x_j; \ux_j+\ut_j/2\bigr)\\
&\hspace{3.7cm}\times\ \sum_{|\alpha'|=n_{c(j)}'}\, F_{\alpha '}(\xi _j)\, \overline{a''}_{\alpha '}\bigl(x_j; \ux_j-\ut_j/2\bigr)\; d\ut_j\period
\end{align*}
In particular, $S_j\in\rrS^{-8-n_j-n_{c(j)}'}(\R^{3(N-1)}\times\R^{3(N-1)})$. Assume that, for some $q>8+n_j+n_{c(j)}'$, $S_j\in\rrS^{-q}(\R^{3(N-1)}\times\R^{3(N-1)})$ then $S_j^0$ must vanish identically. This means that, on a small neighbourhood of $(\hat{\ux}; \hat{\ux}'')$, the above Fourier transform is zero. Since the Fourier transform is injective, 
the product of the two sums must vanish identically as well. Each sum being real analytic, one of them must be zero. By Lemma~\ref{lm:famille-libre}, this implies that either all the functions $a_\alpha$, with $|\alpha|=n_j$, vanish near $\hat{\ux}$ or all the functions $a_{\alpha '}''$, with $|\alpha '|=n_{c(j)}'$, vanish near $\hat{\ux}''$. As in the end of the proof of Lemma~\ref{lm:non-nul}, this contradicts the definition of $n_j$ or the one of $n_{c(j)}'$. \\
By \eqref{eq:décomp-K}, we see that the Weyl symbol of $K$ belongs to $\rrS^{-8-p}(\R^{3(N-1)}\times\R^{3(N-1)})$. Let $\ell\in\cD\setminus\{j\}$. We saw that the nonzero $S_j^0$ and $S_\ell^0$ have different asymptotics as $|\xi|\to\infty$. Therefore any term in the sum in \eqref{eq:décomp-K} cannot be compensated. Since these terms do not belong to $\rrS^{-q}(\R^{3(N-1)}\times\R^{3(N-1)})$, so does the Weyl symbol of $K$. $\cqfd$

\vspace{.5cm}
\begin{appendices}{\bf \Large Appendix.}

\renewcommand{\theequation}{{\rm A}.\arabic{equation}} 

\appendixtitleon
\appendixtitletocon


\vspace{.5cm}

For completeness, we prove in this appendix Lemma~\ref{lm:inté-parties} and Lemma~\ref{lm:famille-libre}.

\Pfof{Lemma~\ref{lm:inté-parties}}Recall that, for the integrable function $g : \R^d\dans\C$, its Fourier transform $F_g$ is given, for $\xi\in\R^d$, by \eqref{eq:fourier-g}, namely 
\begin{equation}\label{eq:fourier-g-bis}
 F_g(\xi)\ =\ \int_{\R^d}\, e^{-i\, \xi\cdot x}\; g(x)\, dx\period
\end{equation}
\begin{enumerate}
 \item Since $g$ is continuous and compactly supported, we can derivate indefinitely many times w.r.t. to $\xi$ through the integral
 in \eqref{eq:fourier-g-bis}. This shows that $F_g$ is smooth. \\
 Since $g$ belongs to the class $\rC^k_c$,
 we can integrate by parts $k$ times in \eqref{eq:fourier-g-bis}, thanks to the identity \eqref{eq:dérivée-exp}.
 This leads to 
\[F_g(\xi)\ =\ |\xi|^{-k}\, \int_{\R^d}\, e^{-i\, \xi\cdot x}\; g_k(x)\, dx\]
 for some compactly supported, continuous function $g_k$. Since the latter integral is bounded w.r.t. $\xi$, we get the desired result.
 \item By assumption, $F$ is integrable on $\R^d$. Thus, we can recover $g$ from $F_g$ by the Fourier inverse formula \eqref{eq:fourier-inv-g-bis}: for $x\in\R^d$, 
\begin{equation}\label{eq:fourier-inv-g-ter}
 g(x)\ =\ (2\pi)^{-d}\int_{\R^d}\, e^{i\, \xi\cdot x}\; F_g(\xi)\, d\xi\period
\end{equation}
By assumption, the partial derivatives of $(x; \xi)\donne e^{i\, \xi\cdot x}\; F_g(\xi)$ w.r.t. $x$ up to order $E(r)$ are $\xi$-integrable thus, by Lebesgue's derivation theorem, we can continuously differentiate $E(r)$ times under the integral sign in \eqref{eq:fourier-inv-g-bis} yielding the $\rC^{E(r)}$ regularity for $g$.$\cqfd$
\end{enumerate}

\Pfof{Lemma~\ref{lm:famille-libre}} Let $n\in\N$. Assume that the map 
\[\R^3\ni\xi\ \donne\ \sum _{|\alpha|=n}\, c_\alpha\, F_\alpha (\xi)\comma\]
for complex coefficients $c_\alpha$, is zero identically. Since the Fourier transform is injective, 
the map 
\begin{equation}\label{eq:sum-chi}
 \R^3\ni x\ \donne\ \sum _{|\alpha|=n}\, c_\alpha\, |x|\, x^\alpha \chi \bigl(|x|\bigr)
\end{equation}
is also zero identically. Thus the real analytic map 
\[\R^3\ni x\ \donne\ \sum _{|\alpha|=n}\, c_\alpha\, x^\alpha \]
is zero identically on the non-empty open set $\{x\in\R^3\setminus\{0\};\, \chi (|x|)=1\}$ and therefore everywhere. Since the homogeneous polynomials $x\donne x^\alpha$, for $|\alpha|=n$, are lineary independent, all the $c_\alpha$ are zero. Thus, the functions $F_\alpha$ are lineary independent. \\
Assume that the map 
\[\R^3\ni\xi\ \donne\ \sum _{|\alpha|=n}\, c_\alpha\, P_\alpha (\xi)\comma\]
for complex coefficients $c_\alpha$, is zero identically. By Lemma~\ref{lm:comportement-F-alpha}, we thus have, for $|\xi|\geq 1$, 
\[\sum _{|\alpha|=n}\, c_\alpha\, F_\alpha (\xi)\ =\ O\bigl(|\xi|^{-5-n}\bigr)\period\]
By Lemma~\ref{lm:inté-parties}, the function \eqref{eq:sum-chi} belongs to $\rC^{n+1}$. By Lemma~\ref{lm:régu-valuation}, this function must be zero identically near $0$. By a previous argument, this shows that all the $c_\alpha$ are zero, yielding the linear independence of the $P_\alpha $, for $|\alpha|=n$. 
$\cqfd$

\end{appendices}

{\bf Data availability and conflict of interest statements.} Data sharing is not applicable to this article as no new data were created or analysed in this study. The authors have no competing interests to declare that are relevant to the content of this article.


%
%
\end{document}